\documentclass[letterpaper,11pt]{article}
\pdfoutput=1 
\usepackage[utf8]{inputenc}

\usepackage[margin=1in]{geometry}
\usepackage{amsfonts}
\usepackage{amsmath}
\usepackage{amssymb}
\usepackage{amsthm}
\usepackage{graphicx}
\usepackage{float}
\usepackage{bbm}
\usepackage[font=small,labelfont=bf]{caption} 
\allowdisplaybreaks

\usepackage{bold-extra} 

\usepackage{hyperref}
\usepackage[svgnames]{xcolor}
\hypersetup{colorlinks={true},urlcolor={blue},linkcolor={DarkBlue},citecolor=[named]{DarkGreen}}

\usepackage{microtype}
\usepackage[capitalise,nameinlink,noabbrev]{cleveref}

\usepackage{xspace}

\usepackage{tikz}
\usepackage{pgfplots}
\pgfplotsset{compat=newest}
\usetikzlibrary{arrows.meta}
\usetikzlibrary{matrix,arrows,shapes,positioning,calc,snakes,decorations.markings,math}
\usetikzlibrary{arrows.meta}
\tikzset{>={Latex[width=5mm,length=5mm]}}

\pgfdeclarelayer{background}
\pgfsetlayers{background,main}

\usetikzlibrary{intersections,decorations.pathreplacing}

\usepackage{doi}

\newcommand{\solnlabel}[1]{\label{sol:#1}}
\newcommand{\solnref}[1]{\ref{sol:#1}}

\DeclareMathOperator{\nextc}{Next}
\DeclareMathOperator{\subline}{Subline}
\DeclareMathOperator{\fix}{Fix}

\theoremstyle{definition}
\newtheorem{definition}{Definition}

\theoremstyle{plain}
\newtheorem{theorem}{Theorem}[section]
\newtheorem{lemma}[theorem]{Lemma}
\newtheorem{corollary}[theorem]{Corollary}
\newtheorem{proposition}[theorem]{Proposition}

\newtheorem{conjecture}[theorem]{Conjecture}

\newtheorem*{lemma*}{Lemma}
\newtheorem*{theorem*}{Theorem}
\newtheorem*{desiredlemma*}{Desired Lemma}

\theoremstyle{remark}

\Crefname{claim}{Claim}{Claims}

\newcommand{\poly}{\textup{poly}}

\def\fp/{\textup{\textsf{FP}}}
\def\p/{\textup{\textsf{P}}}
\def\np/{\textup{\textsf{NP}}}
\def\conp/{\textup{\textsf{co-NP}}}
\def\fnp/{\textup{\textsf{FNP}}}
\def\tfnp/{\textup{\textsf{TFNP}}}
\def\ptfnp/{\textup{\textsf{PTFNP}}}
\def\ppa/{\textup{\textsf{PPA}}}
\def\ppad/{\textup{\textsf{PPAD}}}
\def\ppads/{\textup{\textsf{PPADS}}}
\def\ppp/{\textup{\textsf{PPP}}}
\def\pwpp/{\textup{\textsf{PWPP}}}
\def\pls/{\textup{\textsf{PLS}}}
\def\cls/{\textup{\textsf{CLS}}}
\def\ppadpls/{\textup{$\textsf{PPAD} \cap \textsf{PLS}$}}
\def\ppapls/{\textup{$\textsf{PPA} \cap \textsf{PLS}$}}
\def\eopl/{\textup{\textsf{EOPL}}}
\def\sopl/{\textup{\textsf{SOPL}}}
\def\ueopl/{\textup{\textsf{UEOPL}}}
\def\fixp/{\textup{\textsf{FIXP}}}
\def\bu/{\textup{\textsf{BU}}}
\def\linearfixp/{\textup{\textsf{Linear-FIXP}}}
\def\pspace/{\textup{\textsf{PSPACE}}}

\def\Tarski{\ensuremath{\mathsf{Tarski}}\xspace}

\def\UEOPL{\ensuremath{\mathsf{UEOPL}}\xspace}
\def\SVL{\ensuremath{\mathsf{SVL}}\xspace}
\def\rSVL{\ensuremath{\mathsf{rSVL}}\xspace}
\def\UFEOML{\ensuremath{\mathsf{UFEOML}}\xspace}
\def\OPDC{\ensuremath{\mathsf{OPDC}}\xspace}
\usepackage[inline]{enumitem}

\def\Set#1{\left\{ #1 \right\}}

\newcommand{\eps}{\varepsilon}

\newcommand{\ints}{\mathbb{Z}}

\newcommand{\blank}{\ensuremath{\mathtt{*}}}
\newcommand{\up}{\ensuremath{\mathsf{up}}}
\newcommand{\down}{\ensuremath{\mathsf{down}}}
\newcommand{\zero}{\ensuremath{\mathsf{zero}}}

\title{No Polynomial-Time Algorithm For Tarski Fixed Points Unless Cryptographic Assumptions Fail}

\author{
\begin{tabular}{cc}
& \\
\textbf{Eleni Batziou} & \textbf{John Fearnley}\\
\small{Archimedes/ Athena RC, Greece } & \small{University of Liverpool, UK}\\
\href{mailto:e.batziou@athenarc.gr}{\small{\texttt{e.batziou@athenarc.gr}}} & \href{mailto:john.fearnley@liverpool.ac.uk}{\small{\texttt{john.fearnley@liverpool.ac.uk}}}\\
& \\
\textbf{Alexandros Hollender} & \textbf{Rahul Savani}\\
\small{Technical University of Munich, Germany} & \small{University of Liverpool, UK}\\
\href{mailto:alexandros.hollender@tum.de}{\small{\texttt{alexandros.hollender@tum.de}}} & \href{mailto:rahul.savani@liverpool.ac.uk}{\small{\texttt{rahul.savani@liverpool.ac.uk}}}\\
& \\
\end{tabular}
}

\date{}
 
\begin{document}

\maketitle

\begin{abstract}
We study the time complexity of finding a Tarski fixed point. We provide a
subexponential-time reduction from the Unique End of Potential Line (\UEOPL)
problem to the problem of finding a Tarski fixed point. Using a range of
known cryptographic lower bounds for \UEOPL, our reduction
implies a range of conditional lower bounds on the running time of any
algorithm that finds a Tarski fixed point. The strongest of these give 
quasi-polynomial-time lower bounds for the problem. We also obtain a similar
quasi-polynomial-time lower bound for Tarski based instead on a weak
exponential-time type hypothesis for \UEOPL. These running time lower bounds
match the recently discovered quasi-polynomial query lower bounds for the
problem.
\end{abstract}

\vfill

{\small
\paragraph{\bf AI usage disclosure.} ChatGPT 6-Astra was the primary AI tool
used during the development of this paper. The tool was prompted to investigate
whether an exponential-time hypothesis for \UEOPL implied a lower bound for
\Tarski. It discovered a very complex direct reduction from violation-free
\UEOPL to \Tarski. We then substantially simplified the reduction into the
current exposition, while also extending the results to cover \UEOPL instances
that have violations. All text in the main body of the paper was human
generated, with the AI tool being used for copy-editing purposes only. All
material in the appendix was generated by the tool. 
The proofs of the lemmas presented in Appendices~\ref{app:start} to~\ref{app:tarski-circuits} 
were generated from the human-written statements found in the main
body. The presentation of the cryptographic lower bounds in Appendix~\ref{app:crypto} was generated by the tool following instructions from the authors.
The authors accept all responsibility for the correctness of the results. 
}

\section{Introduction}

Tarski's fixed point theorem~\cite{Tarski55} states that all monotone functions
over a complete lattice have a greatest and least fixed point, and so in
particular, they possess at least one fixed point. In this paper we study the
computational complexity of finding a fixed point in a lattice defined by an
integer grid. A function $f: \{0,1,\dots,n\}^d \rightarrow \{0,1,\dots,n\}^d$
is \textit{monotone} if $x \leq y$ implies that $f(x) \leq f(y)$, where $\le$
denotes the coordinate-wise ordering over the grid. We are
interested in the following 
computational problem called \Tarski: given the function $f$ represented as a Boolean circuit,
find a point $x$ such that $x = f(x)$, or two points $x$ and $y$ that witness a
violation of the monotonicity of $f$. 

The problem is closely connected to the complexity of finding equilibria of
games. Etessami et al. \cite{EPRY20} showed that 
\Tarski is equivalent to finding a pure Nash equilibrium
in a supermodular game. They also showed that the problems of solving
parity games, mean-payoff games and Condon's simple stochastic games reduce to
finding a Tarski fixed point. An efficient algorithm for \Tarski would therefore
have consequences for several longstanding open problems. 

Much of the prior work on \Tarski has considered the query complexity of the
problem, which assumes that $f$ is given as a black box, and measures the
number of queries to $f$ that are needed to find a fixed point, while assuming
that any computation that takes place between queries is free. 
A line of work has produced a series of upper bounds for the query
complexity of \Tarski all of which have query complexity $(\log
n)^{O(d)}$~\cite{DQY11,FPS22,CL22,CLY26}, where successive works have improved
the constant hidden by the $O(d)$.

A recent breakthrough has established a much better algorithm: a
\Tarski fixed point can be found in $O(5^d \cdot (\log n)^{\lceil \log d
\rceil})$ queries~\cite{CGLRY26}. The same paper also gives a lower bound of $\Omega((\log
n)^{\frac{1}{2} \lceil \log d \rceil})$ queries for the problem, implying that
the query complexity of \Tarski is at least quasi-polynomial.

\paragraph{\bf Our contribution.}

We study the \Tarski problem from the point of view of computational
complexity, rather than query complexity. While it is now known that
quasi-polynomially many queries are required to solve the problem when $f$ is
presented as a black box, this does not rule out the existence of a polynomial
time algorithm for the case where $f$ is presented as a Boolean circuit. This
is
because an algorithm might exploit its access to the circuit to circumvent the
lower bound in the query model.

Our main result shows that such an algorithm is unlikely to exist. We show that, if
any of a range of computational or cryptographic assumptions hold, then 
\Tarski cannot be solved in polynomial time. More specifically, we consider the
following two types of assumptions.

\begin{itemize}
\item A weak exponential time hypothesis for the unique end of potential line
problem (\UEOPL), which is the defining problem of the complexity class UEOPL
of the same name. 

\item Cryptographic hardness assumptions that imply sufficiently strong lower bounds for the relaxed Sink of Verifiable Line (\rSVL) problem. \rSVL is the main tool used in the literature to obtain hardness of various TFNP classes from cryptographic assumptions.
\end{itemize}

We show that if any such assumption holds, the \Tarski problem cannot be
solved in polynomial time. Our strongest lower bounds come from the weak ETH
for UEOPL, and, on the cryptographic side, from either (i) repeated squaring
combined with a strong Fiat-Shamir assumption, or alternatively (ii) functional
encryption together with injective one-way functions. Each of these assumptions
individually implies that the \Tarski problem requires quasi-polynomial time,
matching the known query lower bound. 
We discuss each of the lower bounds in more detail in the technical overview in
Section~\ref{sec:techover}.

We prove our result by providing a sub-exponential-time reduction from the
\UEOPL problem to \Tarski, which produces a sub-exponentially sized \Tarski
instance. We show that if an algorithm could solve a \Tarski instance of size
$M$ in time $M^{O((\log M)^{1-\eps})}$ for some $\eps > 0$, then applying it to
our reduced instance would give an algorithm that falsifies the
weak ETH for \UEOPL.

The same reduction also allows us to obtain hardness based on cryptographic assumptions by leveraging the existing reduction from \rSVL to \UEOPL~\cite{FGMS20,HY17}. This means that we can show hardness for \Tarski by making use of existing works that have established hardness of \rSVL from a variety of different cryptographic hardness assumptions. Not all of these give sufficiently strong lower bounds on \rSVL for our purposes, but we show that a number of them do. Our strongest \Tarski lower bound is obtained from the work of Ephraim et al.~\cite{EFKP20}, who proved an almost tight hardness result for \rSVL relying on repeated squaring and strong Fiat-Shamir assumptions. Together with our reduction, this implies that \Tarski cannot be solved in time $M^{o(\log M)}$, unless one of those two cryptographic assumptions fails.

A nice feature of our reduction is that it always produces a monotone \Tarski
instance. In principle, we could prove time lower bounds by always creating a
\Tarski function $f$ that is not monotone and has no fixed point, and providing
a lower bound on the time needed to find a violation of monotonicity. This
would provide a lower bound for \Tarski without saying anything about the
computational complexity of finding a fixed point of a monotone function. Our
result does not do this: even though solutions of a \UEOPL instance are split into
proper solutions and violations, a \UEOPL instance that contains violations
is still mapped to 
a monotone \Tarski instance, and thus we always get a \Tarski instance with no violations.

Technically, our reduction proceeds in two steps. We first reduce \UEOPL to the
one-permutation discrete contraction (\OPDC) problem. There is an existing
reduction between these two problems~\cite{FGMS20}, but we provide a new
reduction that is parameterized by a value $r$ that controls the size of the
resulting \OPDC instance. We then show that the construction used in the query
lower bound for \Tarski~\cite{CGLRY26} can be adapted in order to encode a full
\OPDC problem, including instances that contain violations. A detailed overview
of our techniques can be found in Section~\ref{sec:techovertech}.

\subsection{Related Work}
\paragraph{Query complexity.} 
A long line of work has studied the query complexity of \Tarski, both in terms of upper and lower bounds. Dang, Qi and Ye \cite{DQY11} proposed a recursive binary search algorithm that find a Tarski fixed point in $O((\log n)^d)$ time and queries for fixed dimension $d$. Fearnley, P{\'a}lv{\"o}lgyi and Savani \cite{FPS22} gave a $O(\log^2 n)$ query algorithm for the problem in three dimensions and a decomposition theorem that solves the $d$-dimensional version of the problem in $O(\log^{2 \lceil d/3 \rceil})$ queries. Chen and Li \cite{CL22} improved the bound to $O^{\lceil (d+1)/2 \rceil}$, using the auxiliary problem \textsc{Tarski}*. Haslebacher and Lill \cite{HL26} gave a different optimal algorithm in three dimensions using level sets to obtain the same bound. Chen, Li and Yannakakis \cite{CLY26} obtained an $O(\log^2 n)$ query algorithm in four dimensions, matching the lower bound of \cite{EPRY20}, which extended to an $O(\log^{\lceil (d-1)/3 \rceil + 1} n)$ upper bound via a decomposition theorem for $d$ dimensions. Recent work of Chen et al. \cite{CGLRY26} established the first quasi-FPT upper bound with $O \big( 5^d \cdot (\log n)^{\lceil \log_2 d \rceil} \big)$ queries for $d \geq 3$, thus reducing the exponent from linear to logarithmic in $d$. 

Etessami et al. \cite{EPRY20} proved an $\Omega (\log^2 n)$ randomized query lower bound that holds even for two dimensions and conjectured a lower bound of $\Omega(\log n)^d)$ in general. Br{\^a}nzei, Phillips and Recker \cite{BPR25} improved this to $d$ dimensional instances, proving an $\Tilde{\Omega} (d \log^2 n)$ lower bound, by extending the herringbone construction. Chen et al. \cite{CGLRY26} showed that the exponent of $\log n$ must grow logarithmically with $d$, even for randomized algorithms that always return a fixed point and even when the fixed point is unique. Liu and Wang \cite{LW26} proved a deterministic $\Omega(\log^3 n)$ lower bound in nine dimensions, and Z{\'o}lomy \cite{Zol26} proved the same bound in six dimensions, where it is tight. 

In the quantum query model, Phillips \cite{Phi26} proved an $\Omega(\log^2 n)$ bound in two dimensions and the most recent result of Li et al. \cite{LMYZ26} gave an $\Omega(d \log^2 n)$ lower bound for $d \geq 4$ as well as a superpolynomial lower bound for exponentially wide grids.

\paragraph{Tarski and Total Search Problems.} 
The inclusion of \Tarski in \ppad/ and \pls/ \cite{EPRY20} places the problem in \cls/ by the equality \cls/ = \ppadpls/ \cite{FGHS23}. 
\cite{FGMS20} introduced the class \UEOPL, a restricted subclass of \cls/, which is intended to capture problems with unique solutions, and showed that the unique sink orientation problem, piecewise-linear contraction and P-matrix linear complementarity are contained in this class. 
\cite{CLY23} showed that \Tarski and its restriction to functions with a unique fixed point are equivalent in the black box model. Fearnley and Savani \cite{FS24} showed that the stronger variant, \textsc{Super-Unique-Tarski}, in which every slice has a unique fixed point, is in \UEOPL under promise-preserving reductions.

\paragraph{Cryptographic Lower Bounds.} 
Cryptographic approaches to total search hardness often use the \textsc{Sink of Verifiable Line} (\SVL) problem. In this problem, the goal is to find the last vertex of a line given its starting vertex, a successor circuit and a verifier for the vertex at each position. The verifier makes it possible to embed a sequential computation into a total search problem. This means that one can obtain hardness results for TFNP classes by showing hardness of the \SVL problem. Bitansky, Paneth and Rosen \cite{BPR15} used this approach to obtain average-case hardness for \ppad/ by constructing hard instances of \SVL. Their construction assumes quasi-polynomially secure indistinguishability obfuscation (iO) and subexponential hardness of injective one-way functions. Garg, Pandey and Srinivasan \cite{GPS16} showed that polynomially secure iO and one-way permutations suffice for \ppad/ hardness. Komargodski and Segev \cite{KS17} constructed hard \SVL instances from quasipolynomially secure private-key functional encryption and subexponentially hard injective one-way functions. Hub{\'a}{\v c}ek and Yogev \cite{HY17} obtained cryptographic lower bounds for CLS by reducing \SVL to \textsc{End of Metered Line}. 

Proof systems provide another way to construct verifiable lines. Choudhuri et al. \cite{CHKPRR19a} applied the Fiat-Shamir transformation to the sumcheck protocol to construct instances of relaxed SVL (\rSVL), in which a false positive of the verifier is also a solution. The same authors obtained \ppad/ hardness from iterated squaring in the random oracle model \cite{CHKPRR19b}.

Related work links \ppad/ hardness to  iterated squaring \cite{LV20}, continuous verifiable delay functions \cite{EFKP20} and updatable unambiguous proofs \cite{KPY20}. Jawale et al. \cite{JKKZ21} obtained average-case hardness from subexponential Learning with Errors (LWE) through succinct arguments, and Kalai, Lombardi and Vaikuntanathan \cite{KLV23} obtain hardness from the subexponential Decisional Diffie-Hellman (DDH) assumption.  

Bitansky et al. \cite{BCHKLPR22} derive \ppad/ hardness from polynomial hardness of LWE together with iterated squaring. Chen et al. \cite{CFJW25} construct unambiguous succinct arguments and incrementally verifiably computation from polynomial hardness of LWE. They derive \ppad/ hardness under an additional average-case hardness assumption for polynomial-time functions with a fixed polynomial space bound. Bitansky et al.~\cite{BPSS23} use a related worst-case complexity assumption with fully homomorphic encryption to construct non-interactive universal arguments.

Choudhuri~\cite{Choudhuri26} gave average-case hardness results for \ppad/ based on the Short Integer Solution (SIS) lattice problem, without relying on public-key encryption. Although many of these works state their results for \ppad/, they all actually apply to UEOPL, since both \SVL and \rSVL reduce to \UEOPL.

In a recent result, Fleming et al. \cite{Fleming26} studied TFNP classes in the random oracle model. They proved average-case hardness for \ppad/ and \eopl/, but also gave an algorithm for \ueopl/ that only uses a polynomial number of queries but with unbounded computation between the queries.

\section{Preliminaries}

Our work will use three computational problems, which we now introduce. 

\paragraph{\bf Unique end of potential line.}

The unique end of potential line problem was defined in~\cite{FGMS20}.
Conceptually, it captures a problem in which we are given a single potentially
exponentially long line that is defined implicitly via circuits. Each vertex on
the line is written down using $m$ bits. The line starts at a known vertex
$0^m$, and for every vertex on the line we have a successor circuit $S$ that
gives the next vertex, and a predecessor circuit $P$ that gives the previous
vertex. A directed edge from a vertex $x$ to a vertex $y$ with $x \ne y$ exists
whenever $S(x) = y$ and $P(y) = x$. The initial vertex $0^m$ satisfies $P(0^m)
= 0^m$, and so it has no predecessor. The goal is to find the end of the line,
which is a vertex $x$ such that $P(S(x)) \ne x$, meaning that $x$ has no
outgoing edge. 

The problem also comes equipped with a potential function $V$ that assigns an
$n$-bit potential value to each vertex on the line. It is required that the
potential strictly increases along every edge of the line: if $x \ne y$, $y =
S(x)$, and $x = P(y)$ then we should have $V(x) < V(y)$. 

Of course, many of the restrictions we have described here are not
syntactically enforceable: the circuits $S$, $P$, and $V$ may define multiple
lines, or the potential may decrease along one of those lines. For that reason,
the problem is defined to have proper solutions and violation solutions in the
following way.

\begin{definition}[\UEOPL]
\label{def:UEOPL}
Given Boolean circuits $S,P : \Set{0,1}^m \to \Set{0,1}^m$ such that $P(0^m) =0^m\neq S(0^m)$ and a Boolean circuit $V: \Set{0,1}^m \to \Set{0,1,\dotsc,2^n - 1}$ such that $V(0^m) = 0$, find one of the following:
\begin{enumerate}[label=(U\arabic*), wide=0pt]
\itemsep0mm
\item \solnlabel{U1} A point $x \in \Set{0,1}^m$ such that $P(S(x)) \neq x$.
\end{enumerate}
\begin{enumerate}[label=(UV\arabic*), wide=0pt, leftmargin=\parindent]
\item \solnlabel{UV1} A point $x \in \Set{0,1}^m$ such that $x \neq S(x)$, $P(S(x)) = x$, and $V(S(x)) - V(x) \leq 0$.

\item \solnlabel{UV2} A point $x \in \Set{0,1}^m$ such that $S(P(x)) \neq x \neq 0^m$.

\item \solnlabel{UV3} Two points $x, y \in \Set{0,1}^m$, such that $x \ne y$, $x \ne S(x)$, $y
\ne S(y)$, and either $V(x) = V(y)$ or  $V(x) < V(y) < V(S(x))$.

\end{enumerate}
\end{definition}

Solutions of type \solnref{U1} are the proper solutions that encode the end of
a line. Violations of type \solnref{UV1} consist of an edge on a line at which
the potential does not strictly increase. Violations of type \solnref{UV2} give
us the start of a second line, indicating that the line starting at $0^m$ is
not unique. Violations of type \solnref{UV3} give us another type of evidence
for the existence of multiple lines: if we have two vertices $x$ and $y$ whose
potentials overlap, then we know that either they must belong to distinct
lines, or the single line that contains both $x$ and $y$ does not strictly
increase the potential along every edge. 

We say that an instance is \emph{violation-free} if it has no violation
solutions. It was shown in~\cite{FGMS20} that if a \UEOPL instance is
violation-free, then it has a unique \solnref{U1} solution encoding the end of
the unique line starting at $0^m$. 

Note that an isolated point $x \in \{0, 1\}^m$ with $x = S(x) = P(x)$ can never
be a solution to a \UEOPL instance, and so such points can be ignored. By
contrast, if a point $x \in \{0, 1\}^m$ satisfies $x \ne S(x)$ or $x \ne P(x)$
then it can be a solution, and in a violation-free instance it should be part
of a line. We say that these points are the \emph{vertices} of the instance. 

\paragraph{\bf One-permutation discrete contraction.}

The \emph{one-permutation discrete contraction} problem (\OPDC) will serve as
an intermediary for our reduction. This problem is already known to be
\UEOPL-complete~\cite{FGMS20}, but we will give a strengthened parameterized
reduction from \UEOPL to \OPDC that will be utilized in our eventual reduction
to the problem of finding a Tarski fixed point. 

The \OPDC problem is defined over a discrete grid of points. Let $k = (k_1, k_2,
\dots, k_d)$ be a tuple of dimension widths. We define the $d$-dimensional grid
with widths given by $k$ as $G(k) = [k_1] \times [k_2] \times \dots \times
[k_d]$ where $[n]$ denotes the set $\{0, 1, \dots, n\}$. 

A \emph{slice} of a grid $G = G(k)$ is a $d$-dimensional tuple $s = (s_1, \dots,
s_d)$, where each $s_i \in [k_i] \cup \{\blank\}$. The idea is that 
setting $s_i \ne \blank$ fixes the coordinate to $s_i$, while setting $s_i =
\blank$ leaves dimension $i$ free. Formally, the slice $s$ defines the subset
of points $G_s = \{ x \in G \; : \; x_i = s_i \text{ whenever } s_i \ne \blank \}$. We
will say that a point $x$ lies in the slice $s$ whenever $x$ lies in $G_s$.

An \emph{$i$-slice} is a slice $s$ in which the first $i$ dimensions are
free and the remaining dimensions have been fixed. Formally, this means that
$s_j = \blank$ for all $j \le i$ and $s_j \ne \blank$ for all $j > i$. The
value $i$ in an $i$-slice can range between $0$ and $d$: a $0$-slice contains
only a single point, while a $d$-slice contains the entire grid.

We can now define the \OPDC problem. In this problem we are given a
$d$-dimensional grid of points $G$, and functions $D_1$ through $D_d$ that
are each of the form $D_{i} : G \to \{\up, \down, \zero\}$. It is promised
that the grid has a
global fixed point, which is a point $p$ such that $D_i(p) = \zero$ for
all $i$, and it is also promised that each $i$-slice $s$ has a unique slice
fixed point, which is a point $p \in G_s$ such that $D_j(p) = \zero$ for all $j \le i$.
Finally, it is promised that if $p$ is the fixed point of an $i$-slice $s$ for
$i < d$,
then $D_{i+1}(p)$ tells us the direction to move in dimension $i+1$ in order to find the unique
fixed point of the $(i+1)$-slice $s'$ that contains $s$ (that is, $s'_j = s_j$
for all $j > i+1$). 

The following definition captures this problem, with the violation solutions
capturing violations of the promises defined above. 

\begin{definition}[\OPDC]
\label{def:OPDC}
Given a tuple $k = (k_1, k_2, \dots, k_d)$ defining a grid $G = G(k)$ and circuits $D_i : G \rightarrow
\{\up, \down, \zero\}$ for $i = 1$ through $d$, find one of the following:
\begin{enumerate}[label=(O\arabic*)]
\item \solnlabel{O1} A point $p \in G$ such that $D_i(p) = \zero$ for all $i$.
\end{enumerate}
\begin{enumerate}[label=(OV\arabic*),wide=0pt,leftmargin=\parindent]
\item \solnlabel{OV1} An $i$-slice $s$ and two points $p, q \in G_s$ with $p \ne q$ such that 
 $D_j(p) = D_j(q) = \zero$ for all $j \le i$.
\item \solnlabel{OV2} An $i$-slice $s$, where $i \in \{1, \dots, d\}$, and two points $p, q \in G_s$ such that
\begin{itemize}
\item $D_j(p) = D_j(q) = \zero$ for all $j < i$, 
\item $p_i = q_i + 1$, and
\item $D_i(p) = \down$ and $D_i(q) = \up$.
\end{itemize}
\item \solnlabel{OV3} An $i$-slice $s$, where $i \in \{1, \dots, d\}$, and a point $p \in G_s$ such that
$D_j(p) = \zero$ for all $j < i$, and either
\begin{itemize}
\item $p_i = 0$ and $D_i(p) = \down$, or
\item $p_i = k_i$ and $D_i(p) = \up$.
\end{itemize}
\end{enumerate}
\end{definition}

A solution of type \solnref{O1} is a global fixed point. Violations of type
\solnref{OV1} give two distinct fixed points of an $i$-slice $s$. Violations of
type \solnref{OV2} give two points $p$ and $q$ that lie in the same $i$-slice
$s$, are adjacent in dimension $i$, are fixed points of their respective
$(i-1)$-slices, and point toward each other according to $D_i$. Under the
promise, $D_i$ should point toward the unique fixed point of $s$ at both $p$
and $q$, which cannot be true if $D_i$ points them toward each other. 
Violations of type \solnref{OV3} occur whenever a function $D_i$ points an
$(i-1)$-slice fixed point outside the boundary of the instance, which is a
clear violation of the promise that $D_i$ should point toward the unique fixed
point of the containing $i$-slice.

\paragraph{\bf Tarski.}

Finally we come to the \Tarski problem: find a fixed point of a given monotone
function $f$. 

Given two points $x, y \in G(k)$, we say that $x \preceq y$ if $x_i \le y_i$
for all dimensions $i$. We say that $x \prec y$ if $x \preceq y$ and $x \ne y$. 
A function $f : G(k) \to G(k)$ is \emph{monotone} if for every $x, y \in G(k)$
we have that $x \preceq y$ implies that $f(x) \preceq f(y)$. 
The \Tarski problem is defined as follows.

\begin{definition}[\Tarski]
Given a tuple $k = (k_1, k_2, \dots, k_d)$ defining a grid $G = G(k)$,
and a Boolean circuit defining a function $f : G(k) \rightarrow G(k)$, find one of:
\begin{enumerate}[label=(T), wide=0pt]
\item \solnlabel{T} A point $x \in G(k)$ such that $f(x) = x$.
\end{enumerate}
\begin{enumerate}[label=(TV), wide=0pt]
\item \solnlabel{TV} Two points $x, y \in G(k)$ such that $x \preceq y$ and $f(x) \not\preceq
f(y)$.
\end{enumerate}
\end{definition}

Solutions of type \solnref{T} encode the fixed points of the function $f$,
while violations of type \solnref{TV} encode a violation of the monotonicity of
$f$.

\paragraph{\bf Promise-preserving reductions.}

Let $A$ and $B$ be two problems whose solutions are divided into proper
solutions and violation solutions. 
A reduction from $A$ to $B$ is \emph{promise-preserving} if, when it is
promised that the input instance of $A$ has no violation solutions, then the
resulting instance of $B$ has no violation solutions~\cite{FGMS20}.

\section{Technical Overview}
\label{sec:techover}

Our main theorem is as follows.

\begin{theorem}
\label{thm:main}
Let $\mathcal{L}$ be a \UEOPL instance that can be written down using $N$ bits,
where each vertex uses $m$ bits and each potential value uses $n$ bits with
$n \ge 2$. For every integer $r$ satisfying $1 \le r \le 2^{n-2}$ we can in
time 
$$T = \poly(N, r) \cdot 2^{O(n/\log (r + 1))}$$ 
build a violation-free \Tarski
instance $f : G \to G$ such that $G$ has $2^{O(n/\log (r+1))}$ dimensions that
each have width at most 
$$2^{O(rn(n+m)/\log (r+1))},$$ and every fixed point of
$f$ can be mapped back in time 
$T$ to a
proper solution or violation of $\mathcal{L}$. 
\end{theorem}

The main point to take away from this theorem is that the exponent in the
number of dimensions of the \Tarski instance decreases by a factor of $1/\log(r
+ 1)$ as $r$ increases, while the exponent in the width of the dimensions
increases by a factor of $r / \log(r+1)$. Our lower bounds will be shown by
balancing these two factors. Another important property is that the exponent in
the number of dimensions depends only on $n$, which is the bit-width
of the potentials of $\mathcal{L}$, and it does not depend on $m$, which is the
bit-width of the vertices of $\mathcal{L}$.

\paragraph{\bf A lower bound from an ETH for UEOPL.}

Our main theorem implies a conditional lower bound for \Tarski based on the
following weak exponential-time hypothesis for \UEOPL. 

There is a natural algorithm for the \UEOPL problem that walks along the line
starting at the vertex $0^m$ and stops when it finds a solution or violation.
The length of this walk is limited by the number of distinct potential values
$2^n$, and not by the number of distinct vertex values $2^m$, which could be
much larger. So for a \UEOPL instance of size $N$ with $n$-bit potential values
we get an algorithm that runs in $\poly(N) \cdot 2^{O(n)}$ time.

The weak ETH for \UEOPL conjecture states that we cannot do significantly
better than this for instances where $N \le n^c$ for some constant $c$.

\begin{conjecture}[Weak ETH for \UEOPL]
\label{conj:weak-eth}
The family of \UEOPL instances that can be written down with $N$ bits and that use
$n$-bit potential values, where $N \le n^c$ for some constant $c$, cannot be solved in time
$\poly(N) \cdot 2^{O(n^{1-\eps})}$ for any $\eps > 0$.
\end{conjecture}

Given the recent failure of the ETH for 3SAT~\cite{eth-openai}, it is worth
discussing this conjecture in more detail. 
The first thing to note is that the
failure of the ETH for 3SAT does not directly falsify even the regular ETH for
\UEOPL, which states that there are no algorithms for \UEOPL running in time $\poly(N) \cdot
2^{o(n)}$. There is a linear time reduction from \UEOPL to 3SAT, but applying
it would only give a $\poly(N) \cdot 2^{o(N)}$ algorithm for \UEOPL, and this
does not automatically yield a $\poly(N) \cdot 2^{o(n)}$ algorithm when $N$ can be
polynomially larger than $n$.

Nevertheless, given that there has been progress on ETH-type questions
recently, we have conservatively stated a weaker lower bound conjecture for
\UEOPL, and this conjecture is still good enough to prove a
quasi-polynomial-time lower bound for \Tarski. The following corollary of our
main theorem gives that lower bound.

\begin{corollary}\label{cor:weak-ETH-to-Tarski}
If the Weak ETH for \UEOPL holds, then \Tarski instances that can be written
down in $M$ bits cannot be solved in time $M^{O((\log M)^{1-\eps})}$ for any
$\eps$ satisfying $0 < \eps \le 1$.
\end{corollary}
\begin{proof}
We set $r = 2^{\lfloor \sqrt{n} \rfloor}$.
The reduction in Theorem~\ref{thm:main} with this $r$ value
spends at most $T = \poly(N) \cdot 2^{O(\sqrt{n})}$ time to build 
a \Tarski instance whose encoding size $M$ satisfies $M \le T$, since the size of the
instance cannot exceed the running time of the reduction. 
Moreover, every fixed point of $f$ can be mapped back to a solution of the
original \UEOPL instance in $T$ time.
Observe that $T = 2^{O(\log N)} \cdot 2^{O(\sqrt{n})} = 2^{O(\sqrt{n})}$ due to
the fact that $N \le n^c$.
If the \Tarski instance
could be solved in $M^{O((\log M)^{1-\eps})}$ time for some $0 < \eps \le 1$, then 
we obtain an algorithm for UEOPL that runs in time
\begin{align*}
T + M^{O((\log M)^{1-\eps})} + T &= 2T + 2^{O((\log M)^{2-\eps})} \\
&\le 2T + 2^{O((\sqrt{n})^{2-\eps})} \\
&= 2^{O(\sqrt{n})} + 2^{O(n^{1-\eps/2})} \\
&= 2^{O(n^{1-\eps/2})},
\end{align*}
where the second line uses the fact that $M \le T$, and the final line uses the
fact that $\eps \le 1$. 
This algorithm therefore violates our assumed lower bound on UEOPL.
\end{proof}

We remark that even if the Weak ETH for \UEOPL fails, the proof of this
corollary still implies that there exists an $\eps > 0$ such that \Tarski
cannot be solved in time than $M^{O(\log M)^\eps}$ so long as \UEOPL requires
$\poly(N) \cdot 2^{\Omega(n^{0.5 + \delta})}$ for some constant $\delta > 0$.

\paragraph{\bf Cryptographic lower bounds.}
\UEOPL is known to be hard under a variety of cryptographic hardness assumptions. In fact, almost all of the known cryptographic hardness results for \tfnp/ subclasses already apply to \UEOPL. This is because the vast majority of these results are established for the relaxed Sink of Verifiable Line (\rSVL) problem, and this problem is known to reduce to \UEOPL through a pebbling argument~\cite{BPR15,HY17,FGMS20,Choudhuri26}. Letting $L$ denote the length of the line in an \rSVL instance, and $K$ the number of bits needed to write down the \rSVL instance, our main reduction then yields the following connection:

\begin{quote}
\textit{If the \rSVL problem cannot be solved in time $\poly(K,\log L) 2^{O(\sqrt{\log L})}$, then \Tarski does not admit polynomial-time algorithms.}
\end{quote}

\noindent As a result, existing cryptographic lower bounds for \rSVL that exclude algorithms with such running times immediately imply superpolynomial lower bounds for \Tarski. In fact, some of these cryptographic lower bounds even imply a \emph{quasipolynomial-time} lower bound for \Tarski. These results follow as corollaries by combining our new connection between \Tarski and \rSVL with cryptographic lower bounds in existing work. We summarize these results in Table~\ref{tab:crypto-lower-bounds}. In Appendix~\ref{sec:crypto-efkp} we also show that the result of Ephraim et al.~\cite{EFKP20}, instantiated with appropriate parameters, implies the Weak ETH for \UEOPL (in fact, it rules out $\poly(N)\cdot 2^{o(n)}$-time algorithms), and thus the bound from Corollary~\ref{cor:weak-ETH-to-Tarski}.

The aforementioned connection between \Tarski and \rSVL using our main reduction is formally presented and proved in Appendix~\ref{sec:rsvl-ueopl-tarski}. The corollaries appearing in Table~\ref{tab:crypto-lower-bounds} are presented in Appendices~\ref{sec:crypto-efkp} to \ref{sec:crypto-sis-planted}.

\begin{table}[t]
\centering
\small
\renewcommand{\arraystretch}{1.3}
\begin{tabular}{@{}p{0.13\linewidth}p{0.49\linewidth}p{0.30\linewidth}@{}}
\hline
\textbf{Source} & \textbf{Assumptions} & \textbf{Lower bound for \Tarski} \\
\hline
\cite{EFKP20}
& Repeated squaring and strong Fiat--Shamir
& No $M^{o(\log M)}$-time algorithm \\
\cite{KS17}
& Quasipolynomially secure private-key functional encryption and
sufficiently subexponentially secure injective one-way functions
& No $2^{o((\log M)^{1+\eta})}$-time algorithm, for some constant $\eta>0$ \\
\cite{BCHKLPR22}
& Polynomially hard LWE and iterated squaring
& No polynomial-time algorithm \\
\cite{Choudhuri26}
& Polynomially hard SIS with quasipolynomial norm bound and
exponentially hard planted search
& No polynomial-time algorithm \\
\hline
\end{tabular}
\caption{Cryptographic lower bounds for \Tarski, where $M$ denotes the number of bits needed to write down the instance. All of these lower bounds hold for violation-free \Tarski instances.}
\label{tab:crypto-lower-bounds}
\end{table}

\subsection{Techniques}
\label{sec:techovertech}

We prove the main theorem via a two-step reduction. We first give a
reduction from \UEOPL to \OPDC that is parameterized by the integer $r$, and
then we reduce the \OPDC instance onward to \Tarski. 

\begin{figure}
\centering
\begin{tikzpicture}[
  vertex/.style={draw, circle, minimum width=8mm, minimum height=6mm, inner sep=1pt},
  segment/.style={-{Stealth[length=2mm]}},
  underbrace/.style={decorate, decoration={brace, mirror, amplitude=3pt}}
]
  \node[vertex] (start) at (0,0) {$0^m$};
  \node[vertex] (mid)   at (2.9,0) {$h$};
  \node[vertex] (end)   at (5.8,0) {$e$};

  \begin{scope}[xshift=8cm]
    \node[vertex] (new-start) at (0,0) {$0^m$};
    \node[vertex] (first)     at (2.25,0) {$c_1$};
    \node[vertex] (second)    at (4.5,0) {$c_2$};
    \node[vertex] (new-end)   at (6.75,0) {$e$};
  \end{scope}

  \foreach \source/\target/\i in
    {start/mid/1, mid/end/2, new-start/first/1, first/second/2, second/new-end/3} {
    \draw[segment] (\source.east) -- (\target.west);
    \draw[underbrace] ([yshift=-3mm]\source.east) -- ([yshift=-3mm]\target.west)
      node[midway, below=4pt] {$L_{\i}$};
  }
\end{tikzpicture}

\caption{Left: the existing \UEOPL to \OPDC reduction. Right: our new
reduction.}
\label{fig:lines}
\end{figure}
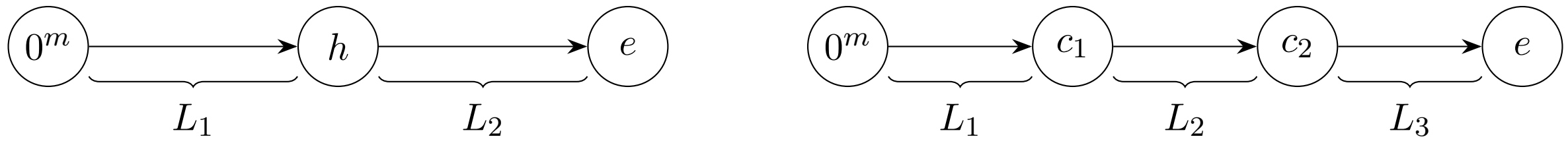

\paragraph{\bf The existing \UEOPL to \OPDC reduction.}

A reduction from \UEOPL to \OPDC is already known~\cite{FGMS20}, and we will
now give a brief summary of it. 

At a high-level, the \OPDC problem can be viewed as encoding a \emph{hierarchical
binary search}. Supposing that the instance is violation-free, we can fix
coordinate $d$ to some value, and then recursively find a fixed point $x$ of the
resulting $(d-1)$-slice. We can then evaluate $D_d(x)$, which gives us a
direction in $\{\up, \down, \zero\}$ to move in dimension $d$. In a
violation-free instance this direction will indeed point toward the location of
the global fixed point in dimension $d$. Thus we can solve an \OPDC instance
using a hierarchy of binary search procedures: we run a binary search in
dimension $d$ where each query requires us to recursively solve a
$(d-1)$-dimensional \OPDC instance. Then each recursive call runs a binary search in
dimension $(d-1)$, and so on.

The existing \UEOPL to \OPDC reduction formulates the \UEOPL problem in a way
that is amenable to hierarchical binary search. The idea is shown on the
left-hand side in Figure~\ref{fig:lines}. Suppose that we have a line that has
exactly $2^n$ vertices for some $n$, and that every step of that line increases
the potential by exactly $1$. The line starts at the vertex $0^m$ that has
potential $0$, and so we are looking for the vertex $e$ that has potential $2^n
- 1$.  

The idea is to use the outer-binary search to guess the vertex $h$ that is
half-way along this line, which specifically means it is the vertex that has
potential $2^{n-1}$. Since we have access to the potential function, we can
easily verify $h$ when we are given it. The vertex $h$ splits the line into two
sub-lines:
\begin{itemize}
\item The line $L_1$ that starts at $0^m$ and ends one step before $h$.
\item The line $L_2$ that starts at $h$ and ends at $e$. 
\end{itemize}
The outer binary search guesses a vertex $x$ for $h$. If the guess is correct then
we recursively solve $L_2$ and find the end of the line $e$. If the guess is
not correct, then we recursively solve $L_1$, which gives us the vertex $y$
that lies one step before $h$. We can then compute $h = S(y)$ using the
successor circuit. Since we now know $h$, we can orient $x$ towards $h$ in the
outer binary search: move $\up$ if the bit encoding of $x$ is below $h$, and
move $\down$ otherwise. Therefore a binary search on the outer dimension can
find $h$. 

The exposition here has assumed that the instance is violation-free. The full
reduction given in~\cite{FGMS20} also handles violations. It also does more
work to make each \OPDC dimension contain only two values, rather than $2^m$ as we
have described it here, but that will not be relevant for our work.

\paragraph{\bf Multiple checkpoints.}

Our new parameterized reduction from \UEOPL to \OPDC follows a similar
approach, but introduces multiple \emph{checkpoints} along the line. That is,
given an integer $1 \le r \le 2^{n-2}$, we distribute $r$ checkpoint vertices
$c_1$, $c_2$, \dots, $c_r$ evenly along the line. The right-hand side of
Figure~\ref{fig:lines} shows the case where $r=2$. The existing reduction can
be thought of as the case where $r = 1$. 

The checkpoints split the line into roughly equal-length segments, and as
before, we can use the potential function to check whether the potential value for a given vertex guess for 
the checkpoint $c_i$ is correct. If we guess vertices for the checkpoints then we
can determine a sub-line that should be recursively solved.
\begin{itemize}
\item If the guess for $c_1$ is incorrect, then we should solve the sub-line $L_1$
starting at $0^m$.
\item If the first $i$ checkpoints are correct, but either the $i+1$th checkpoint is
incorrect or $i = r$, then we should solve the sub-line $L_{i+1}$. 
\end{itemize}
In both cases, recursively solving the sub-line gives us the vertex that lies
one step before the next checkpoint, and we can use that information to fix the
incorrect vertex guess for that checkpoint. In the case where $i = r$ we instead get
the end of the line, which is the vertex that we are looking for.

The key property is that we can still frame this setup as a hierarchical binary
search. We will use the outer binary search to encode the full sequence of
guessed checkpoint vertices $c_1$, $c_2$, \dots, $c_r$, with $c_1$ using the most
significant bits and $c_r$ using the least significant bits. If our guess $x$
for the outer dimension has $c_1$ incorrect, then our recursive call will tell
us the correct vertex for $c_1$, and we can orient $x$ to move $\up$ when the
most significant bits in $x$ are less than $c_1$, and $\down$ when those bits
of $x$ are greater than $c_1$. Once the first checkpoint is correct, the
comparison is then determined by the next incorrect checkpoint, and so on.

We utilize this to produce a reduction from \UEOPL to \OPDC parameterized by
the checkpoint-count $r$. The parameter $r$ gives us a trade-off: increasing
$r$ increases the width of each of the dimensions, since we need wider
dimensions to encode more checkpoints, but then decreases the number of
dimensions, since each recursive call now runs on a shorter line. We will
ultimately use this to obtain our lower bounds for \Tarski.

We have described the technique assuming that the \UEOPL instance is
violation-free. The full reduction also handles \UEOPL instances that contain
violations, and we actually provide a promise-preserving reduction from \UEOPL
to \OPDC.

\paragraph{\bf Reducing \OPDC to \Tarski.}

The recent quasi-polynomial query lower bound for \Tarski~\cite{CGLRY26} can be
viewed as an encoding of an abstract hierarchical binary search into the
\Tarski problem. 
We show that this technique can be extended to encode an \OPDC instance as a violation-free \Tarski instance,  even when the \OPDC instance contains violations.

At a high-level, the reduction given in~\cite{CGLRY26} has three steps. Suppose
that we are encoding an outer binary search on values $1$ through $k$, and that
we have a family of monotone functions $f_1$ through $f_k$ that represent the
inner hierarchical binary searches for each of those values. 

\begin{itemize}
\item The first step is to mirror each of the $f_i$ functions in order to place
the possible locations of a fixed point in an anti-chain. Specifically, 
if $f_i$ is defined over a square grid of width $w$, then they define
$\overline{x} = (w, w, \dots, w) - x$ and set $f'_i(x, y) = (f_i(x),
\overline{f_i(\overline{y})})$. They show that if $f_i$ has a unique fixed
point $x$, then $f'_i$ has a unique fixed point $(x, \overline{x})$. 

\item The second step introduces a new dimension that hides an additional bit
of information for the outer binary search. That is, we can add an additional
monotone dimension to $f'_i$ that hides a $0$ or a $1$ bit at the unique fixed
point $(x, \overline{x})$, and gives no other useful information at any other
point. The mirroring step is what enables us to hide this bit in a monotone
way: since all fixed point candidates $(x, \overline{x})$ lie in an
anti-chain, the information revealed in the extra dimension at the unique fixed
point does not imply, by monotonicity, any information at any of the other
fixed point candidates. 

\item Finally, the first two steps are composed in order to implement the outer
binary search. The second step gives us a way of encoding a single bit with a
monotone extra dimension, but
this is not actually enough to encode a binary search, since the outcome can be $\up$,
$\down$ or $\zero$, which gives three possibilities. This necessitates another
doubling of the dimension so that we can use two monotone extra dimensions to
encode this information. The overall instance $f$ is defined over a tuple
$(a, b, X, Y)$, where $a$ and $b$ are two new dimensions, the $X$ dimensions
encode the mirrored instance $f'_a$, and the $Y$ dimensions encode the mirrored
instance $f'_b$. The key property is that (loosely\footnote{There is an extra
encoding step that places the $f'_i$ functions into the full domain of $f$
that, for the sake of brevity,
we are skipping over here.})
\begin{align*}
f_a(a, b, X, Y) &= b + S^{+}(f'_b(Y)) \\
f_b(a, b, X, Y) &= a - S^{-}(f'_a(X))
\end{align*}
where $S^+$ and $S^-$ are two functions from the second step that hide a bit in
$\{0, 1\}$ that are chosen to encode the outer binary search. 
\end{itemize}

Our contribution is to show that this setup can be extended to encode a full
\OPDC instance. There are several issues that need to be overcome in order to
do this. 

Firstly, the reduction of~\cite{CGLRY26} assumes that each function $f_i$ has a
unique fixed point, but this is not the case in our setting: if the \OPDC
instance has violations, then the resulting \Tarski functions may have
multiple fixed points that map back to violations of the \OPDC instance. This
could break the first step, because the existence of a non-symmetric fixed
point breaks the property that all possible fixed points lie in an anti-chain.
Fortunately, it turns out that each non-symmetric fixed point $(x, \overline{y})$ of the
mirrored instance with $y \ne x$ can be immediately mapped back to a violation
of $\OPDC$, and the fixed points that matter for the binary search remain in an
anti-chain.

Secondly, the existing reduction introduces two new dimensions that each encode
a single bit of information at exactly the symmetrized unique fixed point $(x,
\overline{x})$. In our setting we may have multiple symmetric fixed points, and
so we extend the technique to hide information at multiple symmetric points. In
particular, the idea is that $S^+(x, \overline{x})$ will output $1$ whenever
$D_i(x) = \up$ and $S^-(x, \overline{x})$ will output $1$ whenever $D_i(x) =
\down$, thereby directly encoding the direction function from \OPDC.

Indeed, we show that this encoding gives a very natural mapping back from the
overall fixed points of $f$. The swap between dimensions $a$ and $b$, along
with the fact that $S^+$ and $S^-$ output values in $\{0, 1\}$, ensures that at
any fixed point of $f$ we either have $a = b$ or $a = b + 1$ and $b = a
- 1$. Some of these fixed points will decode to violations in the inner
  instance as described above. Otherwise, the case where $a = b$ will decode to a fixed point of
the \OPDC instance, while in the case where $a = b + 1$ and $b = a - 1$ we will
have two adjacent slice fixed points for which $D_i$ returns $\up$ and $\down$,
respectively, and these will map immediately back to an \solnref{OV2} violation
of the \OPDC instance. 

Finally, it is worth remarking that our extension still keeps the resulting
function monotone, even when the \OPDC instance contains violations. Where
violations from the original instance do appear in $f$, they will appear as
fixed points, so we obtain a violation-free \Tarski instance.

Composing this reduction with our \UEOPL to \OPDC reduction then yields our
main theorem. Note that each step of the reduction blows up the dimension by a
factor of at least $4$, since we mirror the sub-instances, and then include two mirrored
instances in the final function $f$. So the dimension of the resulting instance
is exponential in the number of dimensions in the \OPDC instance. This is where
the parameter $r$ becomes useful, since it allows us to reduce the number of
dimensions in the \OPDC instance, which in turn prevents the resulting \Tarski
instance from becoming too large to prove our lower bounds.

\section{Reducing \UEOPL to \OPDC}

In this section we give a reduction from \UEOPL to \OPDC that is parameterized
by a value $r$, which will be the number of \emph{checkpoints} used in the
construction.

The following lemma allows us to assume that potentials increase by one
along each edge and that every line ends at the same potential.

\begin{lemma}[{\cite[Lemma~17]{FGMS20}}]
\label{lem:2k}
Given a \UEOPL instance $L = (S, P, V)$ such that each vertex of $L$ requires
$m$ bits to write down, and each potential of $L$ requires $n$-bits to write
down, there is a polynomial-time
promise-preserving reduction that produces a \UEOPL instance $L' = (S', P',
V')$ in which each vertex requires $m' = m + n + 1$ bits to write down and each potential
requires $n' = n + 1$ bits to write down, where
\begin{itemize}
\item For every $x$ and $y$ with $x \ne y$, $y = S'(x)$, and $x = P'(y)$, we have $V'(y) = V'(x) + 1$, and
\item We have that $x$ is a solution of type \solnref{U1} in $L'$ if and
only if we have $V'(x) = 2^{n'}-1$.
\end{itemize}
\end{lemma}
The precise bounds on $m'$ and $n'$ in terms of $m$ and $n$ are not explicitly
stated in \cite{FGMS20}, but it is clear from the proof that they hold.
Specifically, every potential used in the reduction is less than or equal to
$2^{n+1} - 1$, and every vertex is a tuple $(x, p)$ where $x$ is a vertex of
the original instance, and $p \le 2^{n'}-1$ is an integer. 

To ease our exposition, we shall describe our construction as if all of the
lines that we consider are violation-free, meaning that they consist of a
single line of length $2^n - 1$. However, our definitions will apply to
instances that have violations, and our proofs will handle those violations. 

\paragraph{\bf Sub-lines.}

Suppose that we have a UEOPL instance $L = (S, P, V)$ such that all
\solnref{U1} solutions $x^*$ of $L$ have $V(x^*) = T$, and for each $x$ and $y$
with $x \ne y$, $y = S(x)$, and $x = P(y)$, we have $V(y) = V(x) + 1$. Note
that an instance that satisfies Lemma~\ref{lem:2k} satisfies this requirement.

Let $x$ be a vertex that lies on a line in $L$ and let $1 \le l < T - V(x)$ be
an integer length. The \emph{sub-line} of $L$ that starts at $x$ and continues for $l$
steps is a \UEOPL instance $L_{x, l} = (S', P', V')$ that is defined as follows.
The circuits $P'$ and $S'$ are defined so that, for each $y \in \{0,1\}^m$,
\begin{align*}
P'(y) &= \begin{cases}
0^m & \text{if $y = 0^m$,} \\
0^m & \text{if $P(y) = x$ and $V(x) < V(y) \le V(x) + l$,} \\
P(y) & \text{if $P(y) \ne x$ and $V(x) < V(y) \le V(x) + l$,} \\
y & \text{otherwise.}
\end{cases} \\
S'(y) &= \begin{cases}
S(x) & \text{if $y = 0^m$,} \\
S(y) & \text{if $y \ne 0^m$ and $V(x) < V(y) \le V(x) + l$,} \\
y & \text{otherwise.}
\end{cases} 
\end{align*}
In other words, the vertex $0^m$, which represents the start of the line, is
identified with $x$ in $L_{x, l}$, and the vertex set of $L_{x, l}$
consists of the vertices within $l$ steps of $x$ along the line, where we
recall that our assumptions imply that each step increases the potential
by exactly $1$. We define the potential function $V'$ so that
\begin{equation*}
V'(y) = \begin{cases}
V(y) - V(x) & \text{if $y \ne x$ and $V(x) \le V(y) \le V(x) + l$,} \\
0 & \text{otherwise.} 
\end{cases}
\end{equation*}
This function simply reduces the potential of the vertices on the line starting
at $x$ by $V(x)$, which puts their potentials into the range $[0, l]$.
Note that this definition also sets $V'(0^m) = 0$, as required.

\begin{lemma}
\label{lem:subline}
Given a \UEOPL instance $L = (S, P, V)$ satisfying the requirements given
above, a vertex $x$, and a length $l$ satisfying $1 \le l < T - V(x)$, there is a polynomial-time algorithm
that computes $L_{x, l}$. Furthermore, the following three properties hold:
\begin{itemize}
\item For every $x$ and $y$ with $x \ne y$, $y = S'(x)$, and $x = P'(y)$, we have
$V'(y) = V'(x) + 1$.
\item If every \solnref{U1} solution of $L$ has potential greater than
$V(x) + l$, then every \solnref{U1} solution $y$ of $L_{x, l}$ has potential $V'(y) = l$. 
\item Every violation of $L_{x, l}$ can be mapped back, in polynomial time, to
a violation of $L$. 
\end{itemize}
\end{lemma}

\paragraph{\bf Checkpoints.}

Suppose that we are given a line $L$ such that every \solnref{U1} solution of
$L$ has potential $T$, where $T \ge 2^4$, and for each $x$ and $y$ with $x \ne
y$, $y = S(x)$, and $x = P(y)$, we have $V(y) = V(x) + 1$. Let $r$ be an
integer between $1$ and $T/4$. We say that a sequence $\langle v_1, v_2, \dots,
v_r\rangle$, where each $v_i \in \{0, 1\}^m$, is a \emph{checkpoint sequence}
for $L$ if all of the following hold for each index $i$.
\begin{itemize}
\item The point $v_i$ is a vertex of $L$, meaning that $v_i \ne S(v_i)$ or $v_i \ne
P(v_i)$.
\item The potential $V(v_i) = \lceil \frac{i}{r+1} \cdot T \rceil$.
\end{itemize}
Note that a checkpoint sequence requires $r \cdot m$ bits to write down.

A \emph{checkpoint candidate} is a sequence $c = \langle v_1, v_2, \dots,
v_r\rangle$, where each $v_i \in \{0, 1\}^m$, which does not necessarily
satisfy the conditions given above. For each checkpoint candidate $c$, we use
$\nextc(c)$ to denote the function that returns the greatest index $i \in \{0,\dots,r\}$ such
that each $v_j$ with $1 \le j \le i$ satisfies the conditions of a checkpoint
sequence. Note that $\nextc(c)$ might be $0$, meaning that the first vertex in
the sequence is incorrect, or $\nextc(c)$ might be $r$, meaning that $c$ is
actually a checkpoint sequence. 

Given a checkpoint candidate $c$ with $i = \nextc(c)$, we define $\subline(c) = L_{v, l}$ where 
$v = v_i$ if $i > 0$, and $v = 0^m$ otherwise, and
$$l = \left\lceil \frac{i+1}{r+1} \cdot T \right\rceil - \left\lceil
\frac{i}{r+1} \cdot T \right\rceil - 1$$
whenever $i < r$, and 
$$l = T - \left\lceil \frac{i}{r+1} \cdot T \right\rceil - 1$$
when $i = r$.
In other words, $\subline(c)$ denotes a sub-line that starts at the last
checkpoint $v_i$ that is correct in $c$, and ends one step before the first
incorrect checkpoint in $c$, or one step before the end of the line $L$ if $c$
is a correct checkpoint sequence. 

Recall that $r \le T/4$ by definition and that $T \ge 2^4$. This implies that
$l \ge 1$, and so by Lemma~\ref{lem:subline} any \solnref{U1} solution $x$ of
$\subline(c)$ cannot have $x = 0^m$, and thus $x$ is in fact some vertex of
$L$.

So if $x$ is a \solnref{U1} solution of $\subline(c)$, and if $i < r$, then we
can use $x$ to fix checkpoint $(i+1)$ in $c$, because $V(x) + 1 = \lceil
\frac{i+1}{r+1} \cdot T \rceil$, and therefore $x$ is the predecessor of a
vertex that can be placed in $v_{i+1}$. So we define the function $\fix(c, x)$
to output the checkpoint candidate $c'$ where $c'_j = c_j$ for all $j \ne
\nextc(c)+1$, and $c'_{\nextc(c)+1} = S(x)$.

\subsection{The reduction to OPDC}

We now reduce the UEOPL instance $L$ to an OPDC instance. 
For the rest of this section we assume that $L = (S, P, V)$ satisfies the
conclusion of Lemma~\ref{lem:2k}. We use $m$ to denote the number of bits
needed to write down a vertex of $L$, and we use $n$ to denote the integer that
defines the potential $2^n - 1$ that all \solnref{U1} solutions must satisfy. 

The reduction will be parameterized by an integer $r$ between $1$ and $2^{n-2}$
that determines the number of checkpoints that will be used in the reduction.

We choose the dimension $d$ of the OPDC instance so that $d = t + 1$, where $t$
is the largest non-negative integer such that $2^n / (r+1)^t \ge 2^4$, or $t = 0$
if there is no such value. 
The widths of the grid $G(k)$ will be defined so that
\begin{itemize}
\item For dimension $1$, we define $k_1 = 2^m - 1$, which allows dimension
$1$ to encode one vertex on the line.
\item For each dimension $i > 1$, we define $k_i =  2^{r \cdot m} - 1$, which
allows dimension $i$ to encode a checkpoint candidate of width $r$.
\end{itemize}

\paragraph{\bf $i$-slice encodings.}

Each $i$-slice of the instance will encode a particular sub-line of $L$, which
we define inductively over the dimensions. If $s$ is an $i$-slice with $i \ge
1$, then we determine the sub-line encoded by $s$ using the following
procedure.
\begin{enumerate}
\item Set $j := d$ and $L' := L$. 
\item If dimension $j$ is free in $s$, then terminate. 
\item Otherwise, let $c =
s_j$, and interpret $c$ as a checkpoint candidate for $L'$. 
Set $L' = \subline(c)$, $j := j - 1$, and return to Step 2.
\end{enumerate}
We define $L_s$ to be the value of $L'$ computed by this procedure for the
slice $s$. 

We also extend this notation to points $x \in G(k)$. For each point $x \in
G(k)$ and index $i$ in the range $1 \le i \le d$, there is a unique $i$-slice
$s$ that contains $x$: we have $s_j = x_j$ for all $j > i$, and $s_j = \blank$
for all $j \le i$. We define the line $L_{x, i} = L_s$.

Before we proceed, we must show that the lines that we construct in this way
satisfy the requirements that we will need. In particular, in the reduction we
will construct checkpoint candidates
with $r$ checkpoints for each line 
$L_{x, i}$ with $i > 1$. The first point of the lemma below states that this is
valid, since each of those lines satisfies the pre-conditions needed for 
checkpoint construction. We will also use the fact that our choice of $d$ ensures
that each line $L_{x, 1}$ is short, which is proved in the second point of the
lemma below. 

The following lemma is proved in Appendix~\ref{app:precondition}.
\begin{lemma}
\label{lem:precondition}
Each line $L_{x, i}$ can be computed in time that is polynomial in the size of
$L$ and $r$, and each violation of $L_{x, i}$ can be mapped back to a violation
of $L$ in time that is polynomial in the size of $L$ and $r$. 
Moreover, the following two properties hold.
\begin{itemize}
\item For every $x \in G(k)$ and $i > 1$ let $L_{x, i} = (S', P', V')$. Every
\solnref{U1} solution of $L_{x, i}$ has potential equal to some value $T$ where $T \ge 16$ and $r
\le T/4$, and for each $p$ and $q$ with $p \ne q$, $q = S'(p)$, and $p =
P'(q)$, we have
$V'(q) = V'(p) + 1$.
\item For every $x \in G(k)$ let $L_{x, 1} = (S', P', V')$. Every \solnref{U1}
solution $p$ of $L_{x, 1}$ satisfies $V'(p) \le 16 \cdot (r + 1)$. 
\end{itemize}
\end{lemma}

\paragraph{\bf An inductive construction.}

We will construct the functions $D_i : G(k) \to \{\up, \down, \zero\}$
inductively over the dimensions. As part of this induction, we maintain the
following invariant: if $x \in G(k)$ satisfies $D_j(x) = \zero$ for all $j \le
i$, then we can in polynomial time compute a solution of $L_{x, i}$ from $x$.

\paragraph{\bf The base case.}

The base case will be the function $D_1 : G(k) \to \{\up, \down, \zero\}$. To
compute $D_1(x)$ we do the following.
\begin{itemize}
\item We consider the line $L_{x, 1}$.

\item By Lemma~\ref{lem:precondition}, the line $L_{x,1}$ has
length at most $16 \cdot (r+1)$. So we
simply follow the line $L_{x,1}$ starting from $0^m$ until we find a point $y$ that
is a
\solnref{U1} solution, or a \solnref{UV1} or \solnref{UV2} violation. 

\item If $x_1 < y$ we set $D_1(x) = \up$, if $x_1 > y$ we set $D_1(x) = \down$, and
if $x_1 = y$ we set $D_1(x) = \zero$. 
\end{itemize}

Observe that $D_1$ can be computed in time that is polynomial in $L$ and $r$.
Specifically, repeated application of Lemma~\ref{lem:subline} implies that
every step that follows the line $L_{x,1}$ must increase the potential by exactly
$1$, and the potential can increase by at most $16 \cdot (r+1)$ before a
\solnref{U1} solution is found. 

Also note that this definition satisfies the invariant: if $D_1(x) = \zero$ for
some point $x \in G(k)$, then $x_1$ encodes a vertex that is a solution of
$L_{x, 1}$.

\paragraph{\bf The inductive step.}

Now suppose that we have defined the functions $D_{1}$ through $D_{i-1}$. We
will define the function $D_{i} : G(k) \to \{\up, \down, \zero\}$ as follows.

Recall that for dimension $i$ we set $k_i = 2^{r \cdot m} - 1$, and thus
dimension $i$ can encode an $r \cdot m$-bit checkpoint candidate. We fix the
encoding so that a candidate $\langle v_1, v_2, \dots, v_r\rangle$ is encoded
so that $v_1$ through $v_r$ are written in order, with $v_1$ using the most
significant bits, and $v_r$ using the least significant bits.

First note that all solutions of the OPDC problem only refer to the output of
$D_i(x)$ when $D_j(x) = \zero$ for all $j < i$. So if $x \in G(k)$ is a point
with $D_j(x) \ne \zero$ for some $j < i$, we arbitrarily set $D_i(x) = \zero$,
and note that this choice cannot affect the reduction.

Now let $x \in G(k)$ be a point such that $D_{j}(x) = \zero$ for all $j < i$. 
\begin{itemize}
\item By the inductive hypothesis, we can compute a solution $y$ of $L_{x, i-1}$
from $x$. If $y$ is a violation, then we set $D_i(x) = \zero$ and
terminate. 

\item Otherwise, recall that we are interpreting $x_i$ as a checkpoint
candidate for $L_{x, i}$, and that $L_{x, i-1} = \subline(x_i)$. So 
the solution $y$ is a \solnref{U1} solution of $\subline(x_i)$.
There are two cases:
\begin{itemize}
\item If $\nextc(x_i) = r$, meaning that $x_i$ is a correct checkpoint
sequence, then $y$ encodes a vertex that is one step before the end of $L_{x,
i}$. We fix $D_i(x) = \zero$ in this case, and note that we can recover the end
of $L_{x, i}$ in polynomial time by taking the successor of the vertex encoded
by $y$.

\item Otherwise, we have 
$\nextc(x_i) < r$ and we can
compute $x'_i = \fix(x_i, y)$. We orient $D_i(x)$ towards $x'_i$: we set
$D_i(x) = \up$ if $x_i < x'_i$ and we set $D_i(x) = \down$ if $x_i > x'_i$.
Here we note that we cannot have $x_i = x'_i$, since $x'_i$ has a correct value
in slot $\nextc(x_i)+1$ of the checkpoint sequence, and $x_i$ does not. 
\end{itemize}
\end{itemize}
Note that this definition satisfies the invariant because we fix $D_i(x) =
\zero$ only when $y$ gives us a violation of $L_{x, i-1}$, which gives us a violation
of $L_{x, i}$ by Lemma~\ref{lem:subline}, or when $y$ gives us a \solnref{U1}
solution of $L_{x, i-1}$, and taking the successor of this vertex gives us a
solution of $L_{x, i}$. 

Finally we must argue that $D_i$ can be computed in polynomial time in the size
of $L$ and $r$. To compute $D_i$ we first need to compute $D_1$ through
$D_{i-1}$ to check that they all return $\zero$, and when they do, to compute
the solution $y$ of $L_{x,i-1}$. Assuming inductively that this can be done in
time that is polynomial in $L$ and $r$, we argue that we can compute $D_i$
in an additional amount of time that is polynomial in $L$ and $r$.
Specifically, computing $D_i$ requires us to evaluate the circuits for
$L_{x,i}$ and $L_{x,i-1}$, and Lemma~\ref{lem:precondition} ensures that these
circuits can be constructed and then evaluated within the required time bound.
All other operations also fit within the required time.

\paragraph{\bf Correctness.}

Our inductive invariant immediately gives us the following lemma, which deals
with the non-violation solutions of the OPDC instance. 

\begin{lemma}
Every \solnref{O1} solution of the OPDC problem can be mapped back, in
polynomial time, to a solution of $L$. 
\end{lemma}
\begin{proof}
Suppose that $x$ satisfies $D_i(x) = \zero$ for all $i$. Then the inductive
invariant says that we can, in polynomial time, compute a solution to $L_{x,
d}$. Since $L_{x, d} = L$, we are done. 
\end{proof}

At an \solnref{OV1} violation, we have some $i$-slice that has two distinct
fixed points $x$ and $y$ according to the $D_j$ functions with $j \le i$. There
are two possibilities here: if a $D_j$ function returned $\zero$ for either of
$x$ or $y$ due to a violation, then this violation can be mapped back to a
violation of the original line via Lemma~\ref{lem:subline}. Otherwise, the
inductive invariant can be used on $x$ and $y$ to find two distinct valid
checkpoint sequences for a sub-instance $L'$ of $L$. We can use these two
sequences to find two distinct vertices in $L'$ that have the same potential.
This gives us a \solnref{UV3} violation of $L'$, and then
Lemma~\ref{lem:subline} can be used to produce a violation of $L$.

The following lemma is proved formally in Appendix~\ref{app:ov1}.
\begin{lemma}
\label{lem:ov1}
Every \solnref{OV1} violation of the OPDC instance can be mapped back, in
polynomial time, to a violation of $L$. 
\end{lemma}

In a \solnref{OV2} violation, we have some $i$-slice that has two points $x$
and $y$ that are adjacent in dimension $i$ and that are fixed points according to
the $D_j$ functions with $j < i$, such that directions chosen by $D_i$ cause
$x$ and $y$ to point toward each other. Since $D_i(x) \ne \zero$ and $D_i(y) \ne
\zero$, we know that no violations were found by the functions $D_j$ with $j <
i$, so we know that $x$ and $y$ decode to two checkpoint candidates for
the same sub-line of $L$. We show that the fact that $x$ and $y$ are
adjacent and point toward each other implies that $D_i$ is attempting to
overwrite the same checkpoint position with two different vertices at $x$ and
$y$. This gives us a sub-line of $L$ that contains two distinct vertices with
the same potential, and this can be translated into a violation of $L$ as
before.

The following lemma is proved formally in Appendix~\ref{app:ov2}.
\begin{lemma}
\label{lem:ov2}
Every \solnref{OV2} violation of the OPDC instance can be mapped back, in
polynomial time, to a violation of $L$. 
\end{lemma}

Finally, we argue that our construction cannot produce an \solnref{OV3}
violation, which consists of a point on the boundary that points outside the
grid according to some $D_i$ function. This cannot occur by 
construction: the function $D_1$ only moves towards some point in $[k_1]$, and
likewise the $D_i$ functions with $i > 1$ only move towards some checkpoint
candidate, and the encodings of all 
checkpoint candidates lie within $[k_i]$. 

The following lemma is proved formally in Appendix~\ref{app:ov3}.
\begin{lemma}
\label{lem:ov3}
The constructed OPDC instance cannot contain an \solnref{OV3} violation.
\end{lemma}

Thus we have shown the following theorem.

\begin{theorem}
\label{thm:opdc}
Given a UEOPL instance $L$ with $m$-bit vertices and $n$-bit potentials that
satisfies the conclusions of Lemma~\ref{lem:2k}, and an integer $r$ between $1$
and $2^{n-2}$, we can, in time polynomial in $r$ and the size of $L$, reduce
$L$ to an OPDC instance with width at most $2^{r \cdot m}$ and $O(n/\log (r+1))$
dimensions. Moreover, the reduction is promise-preserving.
\end{theorem}

\section{Reducing \OPDC to \Tarski}

In this section we show that \OPDC can be reduced to \Tarski. Our construction
closely follows the construction that was used to give a query lower bound for
Tarski~\cite{CGLRY26}. 

We show that an \OPDC instance $\mathcal{O}$ defined by functions $D_1$ through
$D_d$ defined over some grid $G(k)$ can be reduced to the \Tarski instance $f$.
Since the instance provided by Theorem~\ref{thm:opdc} does not have
\solnref{OV3} violations, we will assume throughout this section that
$\mathcal{O}$ also does not have any \solnref{OV3} violations.

\paragraph{\bf The inductive invariant.}

The construction will be inductive over the dimensions: for each $i$-slice $s$,
we will build a \Tarski instance $f_s$ that represents \OPDC instance
restricted to that slice. Moreover, for each index $i$, there will be a single
grid $G_i$ such that the function for each $i$-slice $s$ has the form $f_s :
G_i \to G_i$.

We will maintain the following invariants throughout this process.
\begin{itemize}
\item Each function $f_s$ is monotone.
\item There is a function $m_s$ such
that, for each fixed point $x$ of $f_s$, we have that $m_s(x)$ gives us either
a violation of the \OPDC instance $\mathcal{O}$, or a point $y \in G_s$
that is a fixed point of the slice $s$, meaning that
$D_j(y) = \zero$ for all $j \le i$.
\item The function $m_s$ does not map two distinct fixed points of $f_s$ to the
same fixed point in $s$. The function $m_s$ can, however, map two distinct fixed
points of $f_s$ to the same violation of $\mathcal{O}$.
\end{itemize}

On a practical level, for each $i$, we will actually build a single Boolean
circuit that takes an $i$-slice $s$ and a point $x \in G_i$ and outputs $f_s(x)$,
and another Boolean circuit that takes $s$ and a point $x \in G_i$ that is a
fixed point of $f_s$ and outputs $m_s(x)$.

\paragraph{\bf The base case.}

For the base case, we consider each $1$-slice $s$, where we recall that only
dimension $1$ of a $1$-slice is free, with all others being fixed. We use $x
\circ_1 s$ to denote the operation of fixing dimension $1$ of $s$ to $x$.
Formally, for each $x \in [k_1]$, we have that $y = x \circ_1 s$ is the point in
$G(k)$ with $y_1 = x$ and $y_j = s_j$ for all $j > 1$. 

We build the one-dimensional \Tarski function $f_s : [k_1] \to [k_1]$ as
follows. For each $x \in [k_1]$ we define
\begin{equation*}
f_s(x) = \begin{cases}
x - 1 & \text{if $D_1(x \circ_1 s) = \down$,} \\
x     & \text{if $D_1(x \circ_1 s ) = \zero$, or if $D_1(x \circ_1 s) = \up$ and
$D_1((x+1) \circ_1 s) =
\down$,} \\
x + 1 & \text{otherwise.} \\
\end{cases}
\end{equation*}
For the most part $f_s$ simply moves in the direction specified by $D_1$. The
one exception to this is when $D_1(x \circ_1 s) = \up$ and $D_1((x+1) \circ_1 s) =
\down$, where we force $f_s(x) = x$ instead. This converts an
\solnref{OV2} violation of $\mathcal{O}$ into a fixed point of $f_s$, and
thereby ensures that $f_s$ is monotone.

We can also define the function $m_s$ in the following way. For every $x \in
[k_1]$ such that $x = f_s(x)$ we use the following definitions.
\begin{itemize}
\item The function $m_s(x)$ will return the point $x
\circ_1 s$, which is a fixed point of $s$, in the case where $D_1(x \circ_1 s ) =
\zero$.
\item The function $m_s(x)$ will return the \solnref{OV2} violation defined by $x
\circ_1 s$ and $(x+1) \circ_1 s$ in the case where $D_1(x \circ_1 s) = \up$ and
$D_1((x+1) \circ_1 s) = \down$.
\end{itemize}

It is clear that we can build a Boolean circuit that takes $s$ and $x$ and
outputs $f_s(x)$, and likewise we can build a Boolean circuit that takes $s$
and a point $x$ such that $x = f_s(x)$ and outputs $m_s(x)$. Moreover, the size
of both of these circuits is polynomial in the size of the circuit that
implements~$D_1$. 

The following lemma states that $f_s$ is well-defined and that it satisfies our
inductive invariant. The proof can be found in Appendix~\ref{app:tarski-base}.

\begin{lemma}
\label{lem:tarski-base}
For each $1$-slice $s$, the function $f_s$ defined above does indeed map
$[k_1]$ to $[k_1]$, and $f_s$ is monotone. The function $m_s$ maps every
fixed point of $f_s$ to either a violation of $\mathcal{O}$ or a fixed point of
the slice $s$, and $m_s$ does not map two distinct fixed points of $f_s$ to the
same fixed point of $s$. 
\end{lemma}

\subsection{The inductive step}

Now suppose that for each $(i-1)$-slice $s'$, we have inductively built a
\Tarski instance $f_{s'} : G_{i-1} \to G_{i-1}$ that represents $s'$. We now
build the \Tarski instance $f_s$ for an $i$-slice $s$.

The first step is to transform each instance $f_{s'}$ so that it is defined
over a grid whose width is equal in all dimensions. The following lemma shows
that this can be achieved for any monotone function $f$. Indeed, we can simply
extend each of the smaller dimensions so that they have the same width as the
largest dimension, and then extend the function so that it clamps all points
back into the original grid before applying the original function $f$. The
formal proof of this can be found in Appendix~\ref{app:squaregrid}.
\begin{lemma}
\label{lem:squaregrid}
Given a $d$-dimensional violation-free \Tarski instance $f : G \to G$ defined
over some grid $G$ whose maximum width is $w$, we can in polynomial time
produce a $d$-dimensional violation-free \Tarski instance $f' : G(k') \to
G(k')$ where $k'_i = k'_j = w$ for all $i$ and $j$, such that $G \subseteq
G(k')$ and the fixed points of $f'$ are exactly the same as the fixed points of
$f$. 
\end{lemma}
So from now on, we will assume that every dimension in the grid $G_{i-1}$ has
width $k$. Moreover, the fact that the reduction above does not change the set
of fixed points means that our existing $m_{s'}$ functions for the
$(i-1)$-slices continue to work after the transformation has been applied.

\paragraph{\bf Mirrored instances.}

For each point $x \in G_{i-1}$, we define $\overline{x} = (k, k, \dots, k) - x$
to be the \emph{reflection} of the point $x$. For each $(i-1)$-slice $s'$, we
define the mirrored instance $f'_{s'} : G_{i-1} \times G_{i-1} \to G_{i-1} \times
G_{i-1}$ so that for each $x, y \in G_{i-1}$ we have
\begin{equation*}
f'_{s'}(x, y) = (f_{s'}(x), \overline{f_{s'}(\overline{y})}).
\end{equation*}
This definition matches the one given in~\cite{CGLRY26}. As pointed out in that
work, if $f_{s'}$ is monotone, then $f'_{s'}$ is too. That work assumed that $f_{s'}$
has a unique fixed point $x$, and proved that $(x, \overline{x})$ is the unique
fixed point of $f'_{s'}$. However, in our setting each function $f_{s'}$ will have a
unique fixed point only when the \OPDC instance $\mathcal{O}$ is
violation-free, and $f_{s'}$ may have multiple fixed points when $\mathcal{O}$
contains violations. Here, our inductively defined mapping function $m_{s'}$ will
play a role: if $(x, \overline{y})$ is a fixed point of $f'_{s'}$ with $x \ne
y$, and neither $m_{s'}(x)$ nor $m_{s'}(y)$ return a violation of $\mathcal{O}$, then
$m_{s'}(x)$ and $m_{s'}(y)$ return two distinct fixed points of $s'$, giving us an
\solnref{OV1} violation of $\mathcal{O}$.

\paragraph{\bf Hiding extra information.}

Now we add more dimensions that are designed to hide extra information. The
construction in~\cite{CGLRY26} hides one additional bit of information at the
unique fixed point $(x, \overline{x})$ of $f'_{s'}$ while giving no new
information at any other point. In our setting there can be multiple symmetric
fixed points, so we extend their construction to enable one bit of information
to be hidden at every symmetric point $(x, \overline{x})$.

We first define the two functions that we would like to hide. These functions
are parameterized by a value $t \in [k_i]$, where we recall that $k_i$ is the
width of the $i$th dimension in $\mathcal{O}$. Let $s_t$ be the $(i-1)$-slice
of $\mathcal{O}$ obtained by fixing dimension $i$ in $s$ to $t$. We define the
function $\beta^+_t : G_{i-1} \to \{0, 1\}$ so that, for each $x \in G_{i-1}$,
we have
\begin{equation*}
\beta^+_t(x) = \begin{cases}
1 & \text{if $x = f_{s_t}(x)$, $m_{s_t}(x)$ is not a violation, and $D_i(m_{s_t}(x)) = \up$,} \\
0 & \text{otherwise.} 
\end{cases}
\end{equation*}
We define the function $\beta^-_t : G_{i-1} \to \{0, 1\}$ so that, for each $x
\in G_{i-1}$, we have
\begin{equation*}
\beta^-_t(x) = \begin{cases}
1 & \text{if $x = f_{s_t}(x)$, $m_{s_t}(x)$ is not a violation, and $D_i(m_{s_t}(x)) = \down$,} \\
0 & \text{otherwise.} 
\end{cases}
\end{equation*}
The idea is that these functions encode the directions specified by $D_i$ at
each fixed point of the slice $s_t$: the function $\beta^+_t$ outputs a $1$ whenever $D_i$
outputs $\up$, and $\beta^-_t$ outputs a $1$ whenever $D_i$ outputs $\down$. 
The value of these functions at any point $x$ that is not a fixed point of
$f_{s_t}$ will be irrelevant.

Now we define two corresponding functions that will be used in the construction
of $f_s$. 
Let $C = \{(x, \overline{x}) \; : \; x \in G_{i-1}\}$ be the set of symmetric points
in $G_{i-1} \times G_{i-1}$, and define 
$$D = \{ (a, b) \in G_{i-1} \times G_{i-1} \; : \; \exists x \in G_{i-1} \text{ such that } (a,
b) \preceq (x, \overline{x})\}$$
to be the set of points that lie below the points in $C$. 
For every $t \in [k_i]$ and every $x, y \in G_{i-1}$, we define 
\begin{equation*}
S^+_t(x, y) = \begin{cases}
\beta^+_t(x) & \text{if $(x, y) \in C$,} \\
0 & \text{if $(x, y) \in D \setminus C$,} \\
1 & \text{otherwise.}
\end{cases}
\end{equation*}
We also define
\begin{equation*}
S^-_t(x, y) = \begin{cases}
\beta^-_t(x) & \text{if $(x, y) \in C$,} \\
1 & \text{if $(x, y) \in D \setminus C$,} \\
0 & \text{otherwise.}
\end{cases}
\end{equation*}
Observe that $S^+_t$ is monotone, whereas $S^-_t$ is anti-monotone.

\paragraph{\bf Translating sub-instances.}

The reduction now proceeds along the same lines as the construction given
in~\cite{CGLRY26}. Recall that the width of every dimension in $G_{i-1}$ is
$k$. Let $d$ be the number of dimensions used by $G_{i-1}$. We define $G$ to be
a grid with $2d$ dimensions where each dimension has width $(k_i + 1) \cdot (k+1)$,
where we recall that $k_i$ is the width of dimension $i$ in the OPDC instance
$\mathcal{O}$.

The reduction will require us to translate mirrored instances $f'_{s'}$ for some
$(i-1)$-slice $s'$ to a box $B_t$ in $G$ where $B_t = \{ x \in G \; : \; t \cdot (k
+ 1) \le x_j < (t+1) \cdot (k+1) \text{ for all $j$} \}$. To this end, if
$\mathbbm{1}$ denotes a vector of all ones, then we
define the function $Z_t : G \to G_{i-1} \times G_{i-1}$ so that for each $x \in G$ we have
\begin{equation*}
Z_t(x) = \min \Bigl( \max\bigl( x - t \cdot (k+1) \cdot \mathbbm{1}, 0 \bigr),
k \Bigr),
\end{equation*}
where both the $\min$ and $\max$ operations are applied coordinate-wise. 
That is, the function $Z_t$ maps each point of $B_t$ back to the corresponding
point in $G_{i-1} \times G_{i-1}$, while the $\min$ and $\max$ operations ensure that points
outside $B_t$ are mapped to the boundary of $G_{i-1} \times G_{i-1}$.

\paragraph{\bf The $f_s$ function.}

We can now build the monotone function $f_s$ for our $i$-slice $s$. 
If the grid $G_{i-1}$ used for the $(i-1)$-slice functions has $d$-dimensions,
then $f_s$ will be defined over a grid $G_{i}$ with $4d + 2$ dimensions. 
These dimensions will be split in the following way.
\begin{itemize}
\item The first two dimensions will have width $k_i$, where $k_i$ is the width
of dimension $i$ in $\mathcal{O}$. These dimensions will be used to encode
the $D_i$ function. 
\item The rest of the dimensions will be split into two blocks $X$ and $Y$,
each of which uses $2d$ dimensions.
All of these dimensions will have width $(k_i + 1) \cdot (k+1)$, and each of $X$ and
$Y$ will be used to encode a translated mirrored sub-instance. 
\end{itemize}
Thus, each point in $G_{i}$ can be interpreted as a tuple $(a, b, X, Y)$,
where $a, b \in [k_i]$, and $X$ and $Y$ are points in the grid $G$ defined in
the previous section.

We define the function $f_s : G_i \to G_i$ block-wise in the following way. 
For each $t \in [k_i]$, let $s_t$ be the slice obtained by fixing dimension $i$
of $s$ to $t$.
\begin{align*}
f_s(a, b, X, Y)_X &= a \cdot (k + 1) \cdot \mathbbm{1} + f'_{s_a}(Z_a(X)) \\
f_s(a, b, X, Y)_Y &= b \cdot (k + 1) \cdot \mathbbm{1} + f'_{s_b}(Z_b(Y)) \\
f_s(a, b, X, Y)_a &= \min(k_i, b + S^+_b(Z_b(Y))) \\
f_s(a, b, X, Y)_b &= \max(0, a - S^-_a(Z_a(X)))
\end{align*}
This follows the definition given in~\cite{CGLRY26}, while substituting in our
$S^+_b$ and $S^-_a$ functions that encode the $D_i$ function from
$\mathcal{O}$. In particular, note that the $a$ and $b$ dimensions are swapped
in the sense that their right-hand sides refer to each other, rather than
themselves. This will play an important role in the proof. 

We show that $f_s$ is monotone, and this holds regardless of whether
$\mathcal{O}$ contains violations. At a high-level, this holds because $f_{s'}$
is monotone for each $(i-1)$-slice $s'$ by the inductive hypothesis. The mirroring
operation maintains monotonicity, as does the translation step, so the
dimensions for $X$ and $Y$ are monotone. Dimension $a$ is monotone because
$S^+_b$ is monotone for all $b$, while dimension $b$ is monotone because
$S^-_a$ is anti-monotone for all $a$, and because we subtract this function
rather than adding it. The following lemma is proved in
Appendix~\ref{app:fmonotone}.

\begin{lemma}
\label{lem:fmonotone}
The function $f_s$ does indeed map $G_{i}$ to $G_{i}$ and is monotone. 
\end{lemma}

\paragraph{\bf The $m_s$ function.}

We are also required to produce a function $m_s$ that
maps any fixed point of $f_s$ back to either a violation of $\mathcal{O}$, or
to a fixed point of the slice $s$. We must also ensure that we never map two
distinct fixed points of $f_s$ to the same fixed point of $s$. 

The high-level idea here is that, under the dimension swap between dimensions
$a$ and $b$, at any fixed point of $f_s$ we must either have $a = b$, or $a = b
+ 1$ and $b = a - 1$, because the functions $S^+_b$ and
$S^-_a$ both return values in $\{0, 1\}$. 
These two cases broadly map into two types of solution: if $a = b$ then the
fact that 
$S^+_b$ and $S^-_a$ both return zero implies that $D_i$ cannot
return $\up$ or $\down$ for the grid point encoded at this location, so $D_i$
returns $\zero$, meaning that we have a fixed point of the slice $s$. On the
other hand, if $a = b + 1$ and $b = a - 1$, then we have two points that are
adjacent in dimension $i$ for which $D_i$ returns $\up$ and $\down$,
respectively. This will map directly back to an \solnref{OV2} violation of
$\mathcal{O}$.

There are many details that need to be addressed to make that argument work. In
particular, the solutions to $X$ or $Y$ may not be symmetric, or they may be
symmetric, but the $m_{s_a}$ and $m_{s_b}$ functions map them to a violation of
$\mathcal{O}$. In each of these cases we show that $m_s$ can likewise produce a
violation of $\mathcal{O}$.

Let $(a, b, X,  Y)$ be a fixed point of $f_s$. 
Observe that the $X$ coordinates shift the instance $f'_{s_a}$ up by $a \cdot
(k + 1) \cdot \mathbbm{1}$ and that $Z_a$ undoes exactly the same shift.
Therefore at a fixed point of $f_s$, we have $Z_a(X) = 
f'_{s_a}(Z_a(X))$, which implies that $Z_a(X)$ encodes 
a point $(x, \overline{y}) \in G_{i-1} \times G_{i-1}$ 
that is a fixed point of $f'_{s_a}$. For the same reason, $Z_b(Y)$
encodes a point $(p, \overline{q}) 
 \in G_{i-1} \times G_{i-1}$ that is a fixed point of $f'_{s_b}$. 
\begin{itemize}
\item If $x \ne y$ then 
observe that $x$ and $y$ are two distinct fixed points of $f_{s_a}$.
If either of $m_{s_a}(x)$ or $m_{s_a}(y)$ returns a violation of $\mathcal{O}$,
then $m_s$ can return the same violation.
Otherwise, we have that 
$m_{s_a}(x)$ and $m_{s_a}(y)$ return two distinct fixed points of the slice
$s_a$, which is an \solnref{OV1} violation of $\mathcal{O}$. The function $m_s$
returns this violation.

\item If $p \ne q$ then the same argument can again be used to ensure that
$m_s$ returns a violation of $\mathcal{O}$.

\item The remaining case is where $x = y$ and $p = q$. Here we must consider
the dimensions $a$ and $b$. Since 
$S^+_b$ and $S^-_a$ both return values in $\{0, 1\}$, we either
have $a = b$, or we have $a = b + 1$ and $b = a - 1$. We consider both cases.
\begin{itemize}
\item If $a = b$ then note that $x = f_{s_a}(x)$ and $p = f_{s_b}(p) =
f_{s_a}(p)$. 
\begin{itemize}
\item If
either of $m_{s_a}(x)$ or $m_{s_a}(p)$ return a violation, then $m_s$ will
return the same violation.
\item
If $x \ne p$, then the inductive hypothesis ensures that $m_{s_a}(x)$ and
$m_{s_a}(p)$ are two distinct fixed points of the slice $s_a$, which is an
\solnref{OV1} 
violation of $\mathcal{O}$, which $m_s$ can return. 
\item Finally, we consider the case where $x = p$ and neither of $m_{s_a}(x)$
nor $m_{s_a}(p)$ return a violation. 

We would like to argue that since $a = b$, we have $\beta^+_b(x) =
\beta^-_a(p) = 0$, but we must be careful at the boundary of the instance,
since clipping could cause $a = b$ when $\beta^+_b(x) = 1$ and $b = k_i$ or
$\beta^-_a(p) = 1$ and $a = 0$. But note that both of these cases would give
\solnref{OV3} violations, since $m_{s_a}(x)$ returns a fixed point of the slice
$s_a$, and $D_i(m_{s_a}(x))$ points outside the instance. The \Tarski reduction
has assumed that $\mathcal{O}$ contains no \solnref{OV3} violations, and so we
can indeed proceed with the conclusion that $\beta^+_b(x) = \beta^-_a(p) = 0$.

Since $p = x$, this can only occur when $D_i(m_{s_a}(x)) \ne \up$ and
$D_i(m_{s_a}(x)) \ne \down$, meaning that $D_i(m_{s_a}(x)) = \zero$. So
$m_{s_a}(x)$ is a fixed point of the slice $s$, which $m_s$ can return. 
\end{itemize}

\item If $a = b + 1$ and $b = a - 1$ then we must have $S^+_b(p,
\overline{p}) = 1$, meaning that by definition we have that $p = f_{s_b}(p)$,
that $m_{s_b}(p)$ is not a violation, and that $D_i(m_{s_b}(p)) = \up$.
Likewise we must have $S^-_a(x, \overline{x}) = 1$ implying that $x =
f_{s_a}(x)$, that $m_{s_a}(x)$ is not a violation, and that $D_i(m_{s_a}(x)) =
\down$. Therefore, $m_{s_a}(x)$ and $m_{s_b}(p)$ are an \solnref{OV2} violation
of $\mathcal{O}$, which $m_s$ will return.
\end{itemize}
\end{itemize}

The reasoning above proves the following lemma, which states that $m_s$
satisfies the second property of the inductive hypothesis. 
\begin{lemma}
\label{lem:mcorrect}
For every point $x \in G_i$ such that $x = f_s(x)$ the function $m_s(x)$
produces either a violation of $\mathcal{O}$, or some point $y \in G_s$
satisfying $D_j(y) = \zero$ for all $j \le i$. 
\end{lemma}

We must also show that $m_s$ satisfies the third property of the inductive
hypothesis, which states that $m_s$ cannot map two distinct fixed points of
$f_s$ to the same fixed point of $s$. The following lemma is proved in
Appendix~\ref{app:minjective}.

\begin{lemma}
\label{lem:minjective}
If $x, y \in G_i$ are two distinct points for which $x = f_s(x)$ and $y =
f_s(y)$, and if neither $m_s(x)$ nor $m_s(y)$ return a violation of
$\mathcal{O}$, then $m_s(x) \ne m_s(y)$. 
\end{lemma}

Finally, we note that we can construct a Boolean circuit that, when given an
$i$-slice $s$ and a point $x \in G_i$, computes $f_s(x)$, and likewise we can
construct a Boolean circuit that, when given an $i$-slice $s$ and a point $x \in
G_i$ with $x = f_s(x)$, computes $m_s(x)$.

\paragraph{\bf The overall result for \Tarski.}

Recall that $d$ is the number of dimensions in $\mathcal{O}$. 
We set the final monotone function $f = f_s$ where $s$ is the unique
$d$-slice of $\mathcal{O}$, which is the slice in which $s_j = \blank$ for
all $j$. We will likewise use the function $m = m_s$ to map every fixed point
of $f$ back to a solution of $\mathcal{O}$.

We now compute the number of dimensions used by $f$, and the width of those
dimensions. Let $D_i$ be the number of dimensions used in $G_i$ and let $W_i$
be the maximum width of those dimensions.
\begin{itemize}
\item We have $D_1 = 1$ and $D_i = 4 D_{i-1} + 2$. 
Therefore $D_i \in O(4^i)$.

\item The base case sets $W_1 = k_1$. In the inductive step for dimension $i$
we blow-up the existing dimensions of width $W_{i-1}$ to $(k_i + 1) \cdot (W_{i-1} +
1)$, while introducing two new dimensions of width $k_i$. The transformation
implemented by Lemma~\ref{lem:squaregrid} does not change the width of the
widest dimension.
To bound $W_i$ we can observe that
$$ W_i + 1 \le (k_i + 1) \cdot (W_{i-1} + 1) + 1 \le (k_i + 2) \cdot (W_{i-1} + 1),$$
and therefore 
$$ W_i \le W_i + 1 \le \prod_{j=1}^i (k_j + 2) \le (w + 2)^i.$$
\end{itemize}

Thus, our function $f$ has the form $f : G \to G$ where $G$ is a grid with
$O(4^d)$ dimensions and maximum width at most $(w + 2)^d$. We must also consider the
amount of time that the reduction needs to construct the circuit that computes
$f$, which we do in the following lemma that is proved in Appendix~\ref{app:tarski-circuits}.

\begin{lemma}
\label{lem:tarski-circuits}
Let $N$ be the number of bits used to represent $\mathcal{O}$. The circuits
that compute $f$ and $m$ can be computed in time $\poly(N) \cdot 2^{O(d)}$.
\end{lemma}

The lemmas that we have shown in this section combine to show the following
theorem, which gives the outcome of our reduction from \OPDC to \Tarski.

\begin{theorem}
\label{thm:tarski}
Let $\mathcal{O}$ be an \OPDC instance with $d$ dimensions and maximum width at
most $w$. Let $N$ be the number of bits needed to write down $\mathcal{O}$, and
suppose that $\mathcal{O}$ does not contain a \solnref{OV3} violation. In time
$\poly(N) \cdot 2^{O(d)}$ we can produce a monotone function $f$ defined over a
grid with $O(4^d)$ dimensions that each have width at most $(w+2)^d$ such that
every fixed point of $f$ can be mapped back in time $\poly(N) \cdot 2^{O(d)}$
to a proper solution or violation of $\mathcal{O}$. 
\end{theorem}

\paragraph{\bf Proof of the main theorem.}

We can now combine the reductions that we have specified in the past two
sections to prove our overall main result given in Theorem~\ref{thm:main}. 

Let $\mathcal{L}$ be a \UEOPL instance that can be written down using $N$ bits,
and in which each vertex can be written down in $m$ bits and each potential
value can be written down in $n$ bits where $n \ge 2$. Let $r$ be an integer
satisfying $1 \le r \le 2^{n-2}$.

\begin{itemize}
\item We first apply Lemma~\ref{lem:2k} to produce a second \UEOPL
$\mathcal{L'}$ instance whose encoding uses $\poly(N)$ bits, whose potentials
use $n' = n + 1$ bits, and whose vertices use $m' = m + n + 1$ bits. 

\item We then apply Theorem~\ref{thm:opdc} to $\mathcal{L'}$, which yields an
\OPDC instance $\mathcal{O}$ 
that has $O(n'/\log (r + 1)) = O(n/\log (r + 1))$ dimensions that 
each have width at most $2^{r \cdot m'} = 2^{O(r \cdot (m + n))}$.
The reduction runs in time that is polynomial in $N$ and $r$, and so
$\mathcal{O}$ can be written down in $\poly(N, r)$ many bits. 
Moreover, $\mathcal{O}$ does not contain an \solnref{OV3} violation.

\item Finally we apply Theorem~\ref{thm:tarski} to $\mathcal{O}$ to produce a
\Tarski instance $f$. The reduction and the function that maps solutions of $f$
back both run in time $\poly(N, r) \cdot 2^{O(n/\log (r + 1))}$, and the
resulting \Tarski instance has $2^{O(n/\log (r + 1))}$ dimensions that each
have width at most $2^{O(r n (m + n)/\log (r+1))}$.
\end{itemize}
These three steps produce a violation-free \Tarski instance $f$, and any fixed
point of $f$ can be mapped back to a proper solution or violation of
$\mathcal{L}$.

\bibliographystyle{alphaurl}
\bibliography{references}

\appendix

\section{Proof of Lemma~\ref{lem:subline}}
\label{app:start}

\begin{proof}
Write $a = V(x)$, $b = a+l$, and
\[
I = \{y \in \{0,1\}^m : a < V(y) \le b\}.
\]
By assumption, every \solnref{U1} solution of $L$ has potential $T$, and
every nontrivial reciprocal edge increases the potential by exactly one.
Recall also that $l \ge 1$ and $b < T$.
The circuits $P'$, $S'$, and $V'$ can be constructed in polynomial time from
$S$, $P$, $V$, $x$, and $l$, since their definitions use only evaluations of
the original circuits, comparisons, and subtraction.

We first record a consequence of these assumptions. If $u$ is a
vertex of $L$ with $V(u) \le b$, then $u$ is not a solution of
type~\solnref{U1}, so $P(S(u)) = u$. Moreover, $S(u) \ne u$: otherwise we
would also have $P(u) = u$, making $u$ isolated. The unit-increase property
therefore gives
\begin{equation}
\label{eq:subline-unit-step}
P(S(u)) = u
\qquad\text{and}\qquad
V(S(u)) = V(u)+1.
\end{equation}
In particular, these identities hold for $u=x$.

We next verify that the constructed circuits satisfy the input conditions
of \UEOPL. By definition, $P'(0^m)=0^m$ and $V'(0^m)=0$, and $V'$ takes
values in $\{0,\dots,l\}$. Also, $S'(0^m)=S(x)\ne0^m$. Indeed, if
$S(x)=0^m$, then $x=P(S(x))=P(0^m)=0^m$, contradicting $S(0^m)\ne0^m$.
By \eqref{eq:subline-unit-step}, $V(S(x))=a+1\le b$, so $S(x)\in I$.
The predecessor-redirection rule thus gives
\[
P'(S'(0^m))=P'(S(x))=0^m,
\qquad
V'(S'(0^m))=1.
\]
Hence the new starting point has a consistent outgoing edge that increases
the potential by one.

For the first property, let $u\ne v$ satisfy $v=S'(u)$ and $u=P'(v)$.
If $u=0^m$, the preceding calculation gives $V'(u)=0$ and $V'(v)=1$, as
required. Otherwise, since every nonzero point outside $I$ is isolated in
$L_{x,l}$, we must have $u\in I$. Thus $S'(u)=S(u)=v\ne u$, so $u$ is a
vertex of $L$. By \eqref{eq:subline-unit-step},
\[
P(v)=u
\qquad\text{and}\qquad
V(v)=V(u)+1.
\]
If $V(u)=b$, then $V(v)=b+1$, so $v$ is a nonzero point outside $I$.
It follows that $P'(v)=v$, contradicting $P'(v)=u\ne v$. Hence $V(u)<b$,
and consequently $u,v\in I$. The definition of $V'$ now yields
\[
V'(v)=V(v)-a=V(u)+1-a=V'(u)+1,
\]
which proves the first property.

For the second property, let $y$ be a solution of type~\solnref{U1} in
$L_{x,l}$. Every nonzero point outside $I$ is isolated in $L_{x,l}$, and the
preceding calculation shows that $0^m$ is not a solution of
type~\solnref{U1}. Consequently, $y\in I$. An originally isolated point in
$I$ remains isolated, so $y$ is a vertex of $L$ and
\eqref{eq:subline-unit-step} applies. If $V(y)<b$, then $S(y)\in I$.
Since $P(S(y))=y\ne x$, the predecessor of $S(y)$ is not redirected, and
\[
P'(S'(y))=P'(S(y))=P(S(y))=y,
\]
contradicting the choice of $y$. Therefore $V(y)=b$, and
$V'(y)=V(y)-a=l$, as required.

For the third property, we consider each of the three violation types.
\begin{enumerate}
\item Suppose that $y$ is a violation of type~\solnref{UV1} in $L_{x,l}$.
Since $S'(y)\ne y$, either $y=0^m$ or $y\in I$. The case $y=0^m$ is
impossible, since its outgoing edge increases the potential from zero to
one. If $y\in I$, then $S(y)=S'(y)\ne y$, so
\eqref{eq:subline-unit-step} applies. If $V(y)<b$, its successor remains
in $I$, and
\[
V'(S'(y))=V'(y)+1,
\]
which rules out a violation of type~\solnref{UV1}. If $V(y)=b$, then
$V(S(y))=b+1$, so $S(y)$ is a nonzero point outside $I$ and is isolated in
$L_{x,l}$. Thus
\[
P'(S'(y))=S(y)\ne y,
\]
which also rules out a violation of type~\solnref{UV1}. Hence no violations
of this type occur in $L_{x,l}$.

\item Suppose that $y$ is a violation of type~\solnref{UV2} in $L_{x,l}$,
so $y\ne0^m$ and $S'(P'(y))\ne y$. Since nonzero points outside $I$ are
isolated, $y\in I$. Let $p=P(y)$. If $S(p)\ne y$, then $y$ itself is a
violation of type~\solnref{UV2} in $L$.

Otherwise, $S(p)=y$. We must have $p\ne y$, since $p=y$ would make $y$
isolated in both instances. The original edge from $p$ to $y$ is therefore
reciprocal, and the unit-increase property gives
\[
V(p)=V(y)-1\in[a,b-1].
\]
If $p=x$, then $P'(y)=0^m$ and $S'(0^m)=S(x)=y$, contradicting the
assumed violation. If $V(p)>a$, then $p\in I$, $p\ne x$, and both relevant
pointers are retained, giving $S'(P'(y))=S(p)=y$, again a contradiction.
The remaining possibility is $p\ne x$ and $V(p)=a=V(x)$. Since
$S(p)=y\ne p$ and $S(x)\ne x$, the pair $(p,x)$ is a violation of
type~\solnref{UV3} in $L$.

\item Suppose that $(u,v)$ is a violation of type~\solnref{UV3} in
$L_{x,l}$. In particular, $S'(u)\ne u$ and $S'(v)\ne v$. The analysis of
type~\solnref{UV1} above shows that, for any vertex $z$ with $S'(z)\ne z$,
either
\[
V'(S'(z))=V'(z)+1,
\]
or $z$ lies at the upper cutoff and $V'(S'(z))=0$. Since potentials are
nonnegative integers, neither case permits
$V'(u)<V'(v)<V'(S'(u))$. Thus $V'(u)=V'(v)$.

Neither vertex can be $0^m$: every other vertex with a nontrivial successor
belongs to $I$ and has strictly positive new potential. Hence $u,v\in I$,
where
\[
V'(u)=V(u)-a
\qquad\text{and}\qquad
V'(v)=V(v)-a.
\]
It follows that $V(u)=V(v)$. Their successors are unchanged, so
$S(u)\ne u$ and $S(v)\ne v$. Since $u\ne v$, the same pair $(u,v)$ is
therefore a violation of type~\solnref{UV3} in $L$.
\end{enumerate}
Each mapping uses only a constant number of circuit evaluations and
comparisons, and can therefore be computed in polynomial time.
\end{proof}

\section{Proof of Lemma~\ref{lem:precondition}}
\label{app:precondition}

\begin{proof}
Let $N$ be the encoding size of $L$, and write $q=r+1$. Fix $x\in G(k)$,
and let
\[
L^{(j)}=L_{x,d-j}\qquad(0\le j\le t)
\]
be the sequence of instances obtained in the slice-encoding procedure.
In particular, $L^{(0)}=L$ and $L^{(t)}=L_{x,1}$.

We first bound the computational cost of the sub-line constructions. We
verify below that each construction satisfies the preconditions of
Lemma~\ref{lem:subline}. Write
\[
L^{(j)}=(S_j,P_j,V_j),
\]
and let $\sigma_j,\pi_j,\nu_j$ denote the respective circuit sizes. All
instances use $m$-bit vertices, and their potentials can be represented
using $n$ bits.

Consider a sub-line construction taking $L^{(j)}$ to $L^{(j+1)}$, with
starting vertex $v$ and length $l$. Once $v$ and $l$ have been determined,
we evaluate $V_j(v)$ and $S_j(v)$ and insert their values as constants into
the new circuits. The definition of $V_{j+1}$ then requires one copy of
$V_j$, together with comparisons and subtraction. The definition of
$S_{j+1}$ requires one copy of $S_j$ and one copy of $V_j$, together with
comparisons and selection operations. Likewise, $P_{j+1}$ requires one
copy of $P_j$ and one copy of $V_j$. Whenever a circuit output is used
more than once, the corresponding gates are shared.

Thus there is a fixed polynomial $p$ such that
\[
\begin{aligned}
\nu_{j+1}&\le\nu_j+p(N),\\
\sigma_{j+1}&\le\sigma_j+\nu_j+p(N),\\
\pi_{j+1}&\le\pi_j+\nu_j+p(N).
\end{aligned}
\]
The initial circuit sizes are at most $N$. It follows that
\[
\nu_j\le N+jp(N)
\]
and
\[
\sigma_j,\pi_j\le(j+1)N+\frac{j(j+1)}2p(N).
\]
Since $q\ge2$, the nesting depth satisfies $t\le n$, so all these
circuits have size polynomial in $N$.

At each level, the value of $\nextc$ can be found by checking at most $r$
checkpoint vertices. Each check uses evaluations of the current
successor, predecessor, and potential circuits, together with arithmetic
on polynomially many bits. The terminal potential is initially $2^n-1$
and is updated to the chosen sub-line length after each construction.
Thus the starting vertex, sub-line length, and constants needed for the
next construction can all be computed in time polynomial in $N$ and $r$.
There are at most $t\le n$ levels, and generating each new circuit takes
time polynomial in its size. Consequently, every $L_{x,i}$ can be
constructed in time polynomial in $N$ and $r$.

Now suppose that a violation of $L_{x,i}=L^{(d-i)}$ is given. Retain the
intermediate circuits and the starting vertices and lengths used in
their construction. Applying the violation map from
Lemma~\ref{lem:subline} successively gives a violation of
$L^{(d-i-1)}$, then of $L^{(d-i-2)}$, and so on, until a violation of
$L^{(0)}=L$ is obtained. Each application uses only a constant number of
evaluations of the relevant circuits and comparisons. The circuits have
polynomial size, every intermediate violation consists of at most two
$m$-bit vertices, and there are at most $t\le n$ applications. Including
construction of the intermediate instances, the total running time is
polynomial in $N$ and $r$.

It remains to verify that the sub-line constructions satisfy the required
preconditions and establish the two numbered properties. We show
inductively that each $L^{(j)}$ is well-defined, that all its
solutions of type~\solnref{U1} have a common potential $T_j$, and that
every nontrivial reciprocal edge increases the potential by one. We also
establish the bounds
\begin{equation}
\label{eq:precondition-bounds}
\frac{2^n}{q^j}-4<T_j<\frac{2^n}{q^j}.
\end{equation}
For $j=0$, the required properties follow from the normalization of $L$,
with $T_0=2^n-1$, which satisfies \eqref{eq:precondition-bounds}.

Now suppose that these properties hold at some level $j<t$. Since a
further subdivision is required, the definition of $t$ gives
\[
\frac{2^n}{q^j}\ge16q.
\]
Consequently,
\[
T_j>\frac{2^n}{q^j}-4\ge16q-4.
\]
As $q=r+1\ge2$, this implies both $T_j\ge16$ and $r\le T_j/4$.
Thus $L^{(j)}$ satisfies the preconditions of the checkpoint construction.

Let $a\in\{0,\dots,r\}$ be the value of $\nextc$ for the checkpoint
candidate used at this level. The length of the resulting sub-line is
\[
l=\left\lceil\frac{(a+1)T_j}{q}\right\rceil
-\left\lceil\frac{aT_j}{q}\right\rceil-1.
\]
This formula also covers $a=r$, since $q=r+1$. The checkpoint
preconditions ensure that $l\ge1$, and the upper cutoff has potential
\[
\left\lceil\frac{(a+1)T_j}{q}\right\rceil-1<T_j.
\]
Lemma~\ref{lem:subline} therefore applies: the resulting instance
$L^{(j+1)}$ has common terminal potential $T_{j+1}=l$, and its
nontrivial reciprocal edges increase the potential by one.

The difference of the two ceilings is either $\lfloor T_j/q\rfloor$ or
$\lceil T_j/q\rceil$. Hence
\[
\left\lfloor\frac{T_j}{q}\right\rfloor-1
\le T_{j+1}\le
\left\lceil\frac{T_j}{q}\right\rceil-1,
\]
and in particular
\[
\frac{T_j}{q}-2<T_{j+1}<\frac{T_j}{q}.
\]
Using the induction hypothesis and $q\ge2$, we obtain
\[
T_{j+1}>\frac{2^n}{q^{j+1}}-\frac4q-2
\ge\frac{2^n}{q^{j+1}}-4,
\]
while
\[
T_{j+1}<\frac{T_j}{q}<\frac{2^n}{q^{j+1}}.
\]
This proves \eqref{eq:precondition-bounds}, completes the induction, and
verifies the preconditions used above.

For the first property of the lemma, any $L_{x,i}$ with $i>1$
corresponds to a level $j=d-i<t$. The preceding argument gives a common
terminal potential $T_j\ge16$ with $r\le T_j/4$, together with the
required unit-increase property.

For the second property, the choice of $t$ implies
\[
\frac{2^n}{q^t}<16q.
\]
Indeed, when $2^n\ge16$, this follows from the maximality of $t$; when
$2^n<16$, we have $t=0$ and the inequality holds directly. Applying
\eqref{eq:precondition-bounds} at the final level gives
\[
T_t<\frac{2^n}{q^t}<16q=16(r+1).
\]
Every solution of type~\solnref{U1} of $L_{x,1}$ therefore has potential
at most $16(r+1)$, as required.
\end{proof}

\section{Proof of Lemma~\ref{lem:ov1}}
\label{app:ov1}

\begin{proof}
Let $s$ be an $i$-slice and let $p\ne q$ be points in $G_s$ satisfying
\[
D_j(p)=D_j(q)=\zero
\qquad\text{for all }j\le i.
\]
Let $h$ be the largest index such that $p_h\ne q_h$. Since $p$ and $q$
belong to the same $i$-slice, we have $h\le i$. Moreover, they agree in
every coordinate greater than $h$, so
\[
L_{p,h}=L_{q,h}.
\]
Write $M=(S_M,P_M,V_M)$ for this common instance.

First, $h\ne1$. Indeed, if $h=1$, then the base-case procedure follows the
same instance $L_{p,1}=L_{q,1}$ and finds the same solution $y$ when
evaluating either point. The equalities $D_1(p)=D_1(q)=\zero$ therefore
imply $p_1=y=q_1$, contradicting the choice of $h$.

Thus $h\ge2$. Run the decoding procedures used to evaluate $D_h(p)$ and
$D_h(q)$, obtaining solutions of
\[
L_{p,h-1}=\subline_M(p_h)
\qquad\text{and}\qquad
L_{q,h-1}=\subline_M(q_h),
\]
where the subscript $M$ indicates that the checkpoint candidates are
interpreted with respect to $M$. These procedures apply because all the
lower-indexed directions vanish at both points. If either decoded solution
is a violation, Lemma~\ref{lem:subline} maps it to a violation of $M$.

Otherwise, both decoded solutions are of type~\solnref{U1}. Since
$D_h(p)=D_h(q)=\zero$, the definition of $D_h$ implies
\[
\nextc_M(p_h)=\nextc_M(q_h)=r.
\]
Consequently, $p_h$ and $q_h$ encode valid checkpoint sequences for $M$.
They are distinct, so there is some checkpoint position $a$ at which their
vertices $u$ and $v$ differ. If $T_M$ denotes the common terminal potential
of $M$, validity gives
\[
u\ne v,
\qquad
V_M(u)=V_M(v)=\left\lceil\frac{aT_M}{r+1}\right\rceil<T_M,
\]
where the strict inequality follows from $a\le r$ and $r\le T_M/4$.

Both $u$ and $v$ are vertices of $M$, and neither is a solution of
type~\solnref{U1} because its potential is below $T_M$. Hence
\[
P_M(S_M(u))=u,
\qquad
P_M(S_M(v))=v.
\]
In particular, $S_M(u)\ne u$: otherwise the first equality would also give
$P_M(u)=u$, contradicting that $u$ is a vertex. The same argument gives
$S_M(v)\ne v$. Thus $(u,v)$ is a violation of type~\solnref{UV3} in $M$.

In either case, we obtain a violation of $M$. Repeated application of
Lemma~\ref{lem:subline} along the sequence of sub-line constructions
defining $M$ maps this violation to a violation of $L$. The decoding
procedures, comparison of the checkpoint sequences, and at most $d-1$
applications of the violation map take polynomial time.
\end{proof}

\section{Proof of Lemma~\ref{lem:ov2}}
\label{app:ov2}

\begin{proof}
Let $s$ be an $i$-slice and let $p,q\in G_s$ witness a violation of
type~\solnref{OV2}. Thus
\[
D_j(p)=D_j(q)=\zero\quad\text{for all }j<i,
\qquad p_i=q_i+1,
\]
and
\[
D_i(p)=\down,\qquad D_i(q)=\up.
\]
Since $p$ and $q$ agree in every coordinate greater than $i$, they determine
the same instance
\[
M=L_{p,i}=L_{q,i}.
\]
Write $S_M$, $P_M$, and $V_M$ for its circuits.

First, $i\ne1$. Otherwise, the base-case procedure finds the same solution
$y$ in $M$ when evaluating either point. The two directions would imply
\[
q_1<y<p_1=q_1+1,
\]
which is impossible because $y$ is an integer.

Thus $i\ge2$. Write
\[
c=\langle c_1,\dots,c_r\rangle,
\qquad e=\langle e_1,\dots,e_r\rangle
\]
for the checkpoint candidates encoded by $q_i$ and $p_i$, respectively.
Run the decoding procedures used to evaluate $D_i(q)$ and $D_i(p)$.
Since both directions are nonzero, neither procedure returns a violation,
and neither candidate is a valid checkpoint sequence. Instead, each
procedure obtains a solution of type~\solnref{U1} of its corresponding
sub-line and uses it to repair the first incorrect checkpoint.

Let
\[
j=\nextc_M(c)+1,\qquad k=\nextc_M(e)+1,
\]
where the subscript $M$ indicates that the candidates are interpreted with
respect to $M$. Let $u$ and $v$ be the valid checkpoints inserted at
positions $j$ and $k$, respectively. Since each repair changes only one
block, the directions give
\[
u>c_j,\qquad v<e_k.
\]
We will show that $j=k$ and $u\ne v$.

Let $a$ be the first checkpoint position at which $c$ and $e$ differ, and
put $B=2^m$. Because their integer encodings are consecutive and the first
checkpoint occupies the most significant block,
\[
c_b=e_b\quad(b<a),\qquad e_a=c_a+1,
\]
and
\[
c_b=B-1,\qquad e_b=0\quad(b>a).
\]
There are two cases.
\begin{itemize}
\item Suppose that the common prefix contains an incorrect checkpoint.
Both candidates then have the same first incorrect position, so $j=k<a$.
Moreover, $c_j=e_j$, and hence
\[
u>c_j=e_j>v.
\]
Thus $u\ne v$.

\item Suppose that every checkpoint in the common prefix is correct.
Then $j,k\ge a$. We cannot have $j>a$, because then $c_j=B-1$, whereas
the upward direction requires $u>c_j$. Similarly, we cannot have $k>a$,
because then $e_k=0$, whereas the downward direction requires $v<e_k$.
Consequently, $j=k=a$. Since $e_a=c_a+1$, integrality gives
\[
u\ge c_a+1=e_a>c_a\ge v.
\]
Again, $u\ne v$.
\end{itemize}

In either case, $u$ and $v$ are distinct valid checkpoints at the same
position $j$ in $M$. Writing $T_M$ for its terminal potential, we therefore
have
\[
V_M(u)=V_M(v)=\left\lceil\frac{jT_M}{r+1}\right\rceil<T_M,
\]
where the strict inequality follows from $j\le r$ and $r\le T_M/4$.
Both are vertices, and neither is a solution of type~\solnref{U1} because
its potential is below $T_M$. Therefore
\[
P_M(S_M(u))=u,\qquad P_M(S_M(v))=v.
\]
These identities imply $S_M(u)\ne u$ and $S_M(v)\ne v$: otherwise the
corresponding vertex would be isolated. Thus $(u,v)$ is a violation of
type~\solnref{UV3} in $M$.

Finally, repeated application of Lemma~\ref{lem:subline} maps this
violation through the sub-line constructions defining $M$ to a violation
of $L$. The decoding procedures, comparison of the checkpoint candidates,
and at most $d-1$ applications of the violation map take polynomial time.
\end{proof}

\section{Proof of Lemma~\ref{lem:ov3}}
\label{app:ov3}

\begin{proof}
Let $x\in G(k)$ and $i\in\{1,\dots,d\}$. We show that
\[
x_i=0\implies D_i(x)\ne\down,
\qquad
x_i=k_i\implies D_i(x)\ne\up.
\]

If $i=1$, the base-case procedure chooses a vertex $y$, whose integer
encoding belongs to $[k_1]$, and directs $D_1(x)$ towards $y$. Thus
$D_1(x)=\down$ requires $x_1>y\ge0$, which is impossible when $x_1=0$.
Similarly, $D_1(x)=\up$ requires $x_1<y\le k_1$, which is impossible
when $x_1=k_1$.

Now suppose that $i>1$. If $D_i(x)=\zero$, neither forbidden boundary
direction occurs. Otherwise, the construction computes a repaired
checkpoint candidate $x'_i=\fix(x_i,y)$ and directs $D_i(x)$ towards
$x'_i$. This candidate consists of $r$ entries represented by $m$ bits
each, so
\[
0\le x'_i\le 2^{rm}-1=k_i.
\]
Consequently, $D_i(x)=\down$ requires $x_i>x'_i\ge0$, while
$D_i(x)=\up$ requires $x_i<x'_i\le k_i$. The former is impossible at
$x_i=0$, and the latter is impossible at $x_i=k_i$.

Hence no direction points out of the grid, and the constructed instance
contains no violation of type~\solnref{OV3}.
\end{proof}

\section{Proof of Lemma~\ref{lem:tarski-base}}
\label{app:tarski-base}

\begin{proof}
Fix a $1$-slice $s$, and write
\[
v_x=x\circ_1 s
\]
for each $x\in[k_1]$. Since $s$ is a $1$-slice, every point $v_x$ belongs
to $G_s$.

We first show that $f_s(x)\in[k_1]$ for every $x\in[k_1]$. By definition,
\[
f_s(x)\in\{x-1,x,x+1\},
\]
so only the boundary points require consideration. Since $\mathcal{O}$ has
no \solnref{OV3} violations, we have
\[
D_1(v_0)\ne\down
\qquad\text{and}\qquad
D_1(v_{k_1})\ne\up.
\]
Thus $f_s(0)\ne-1$. At the upper boundary, $D_1(v_{k_1})$ is either
$\zero$ or $\down$, so $f_s(k_1)\le k_1$. Hence $f_s$ maps $[k_1]$ to
itself. Moreover, whenever the definition requires evaluating
$D_1(v_{x+1})$, we have $D_1(v_x)=\up$, which implies $x<k_1$; this
evaluation is therefore within the grid.

To prove monotonicity, it suffices to show that
\[
f_s(x)\le f_s(x+1)
\]
for every $0\le x<k_1$. The definition gives
\[
f_s(x)\le x+1
\qquad\text{and}\qquad
f_s(x+1)\ge x.
\]
Since these values are integers, the only way that $f_s(x)>f_s(x+1)$
could hold is if
\[
f_s(x)=x+1
\qquad\text{and}\qquad
f_s(x+1)=x.
\]
The second equality implies $D_1(v_{x+1})=\down$. The first equality
implies $D_1(v_x)=\up$, but the definition of $f_s$ assigns $f_s(x)=x$
whenever
\[
D_1(v_x)=\up
\qquad\text{and}\qquad
D_1(v_{x+1})=\down.
\]
This is a contradiction. Consequently, $f_s$ is monotone.

Now let $x$ be a fixed point of $f_s$. By the definition of $f_s$, exactly
one of the following cases holds.
\begin{itemize}
\item If $D_1(v_x)=\zero$, then $v_x$ is a fixed point of the slice $s$,
and $m_s(x)$ returns $v_x$.
\item Otherwise,
\[
D_1(v_x)=\up
\qquad\text{and}\qquad
D_1(v_{x+1})=\down.
\]
Both points belong to $G_s$, and their first coordinates differ by one.
There are no lower-indexed direction conditions to check when $i=1$.
Thus, taking $p=v_{x+1}$ and $q=v_x$, these points form an
\solnref{OV2} violation, which $m_s(x)$ returns.
\end{itemize}
Therefore every fixed point of $f_s$ is mapped to either a violation of
$\mathcal{O}$ or a fixed point of $s$.

Finally, suppose that $x$ and $y$ are fixed points of $f_s$ whose decoded
outputs are both proper fixed points of $s$. Then
\[
m_s(x)=v_x
\qquad\text{and}\qquad
m_s(y)=v_y.
\]
If $m_s(x)=m_s(y)$, equality of their first coordinates gives $x=y$.
Hence $m_s$ cannot map two distinct fixed points of $f_s$ to the same
fixed point of $s$.
\end{proof}

\section{Proof of Lemma~\ref{lem:squaregrid}}
\label{app:squaregrid}

\begin{proof}
Write
\[
G=[k_1]\times\cdots\times[k_d],
\qquad
w=\max_{1\le j\le d}k_j,
\]
and let $G'=[w]^d$. Thus $G\subseteq G'$, and every dimension of $G'$
has width $w$.

Define $\pi:G'\to G$ by
\[
\pi(x)_j=\min\{x_j,k_j\}
\]
for each $j\in\{1,\dots,d\}$, and define $f':G'\to G'$ by
\[
f'(x)=f(\pi(x)).
\]
This is well-defined because $\pi(x)\in G$ and
$f(\pi(x))\in G\subseteq G'$.

We first show that $f'$ is monotone. Let $x,y\in G'$ satisfy
$x\preceq y$. For every coordinate $j$, we have
\[
\min\{x_j,k_j\}\le\min\{y_j,k_j\},
\]
so $\pi(x)\preceq\pi(y)$. Since $f$ is monotone, it follows that
\[
f'(x)=f(\pi(x))
\preceq f(\pi(y))=f'(y).
\]
Hence $f'$ is monotone, and the resulting \Tarski instance is
violation-free.

We next show that the fixed points of $f'$ are exactly those of $f$.
If $x\in G$ is a fixed point of $f$, then $\pi(x)=x$, and therefore
\[
f'(x)=f(\pi(x))=f(x)=x.
\]
Conversely, suppose that $x\in G'$ is a fixed point of $f'$. Since every
output of $f'$ belongs to $G$, the equality $x=f'(x)$ implies that
$x\in G$. Consequently, $\pi(x)=x$, and
\[
f(x)=f(\pi(x))=f'(x)=x.
\]
Thus the two functions have exactly the same fixed points.

Finally, a circuit for $\pi$ can be constructed using a comparison and
a selection operation in each coordinate. These operations have size
polynomial in the binary encoding length of the grid widths. Composing
this circuit with the given circuit for $f$ produces a circuit for $f'$
in polynomial time.
\end{proof}

\section{Proof of Lemma~\ref{lem:fmonotone}}
\label{app:fmonotone}

\begin{proof}
Fix an $i$-slice $s$. By the inductive hypothesis and
Lemma~\ref{lem:squaregrid}, each child function $f_{s_t}$ is a monotone
map from $[k]^d$ to itself. Write
\[
h=k+1.
\]

We first record the monotonicity properties of the functions used in the
construction. Reflection reverses the coordinate-wise order, so the function
\[
y\longmapsto\overline{f_{s_t}(\overline y)}
\]
is monotone. Consequently, the mirrored function $f'_{s_t}$ is monotone.
For each fixed $t$, the map $Z_t$ is also monotone, since translation and
coordinate-wise clipping preserve the order.

We next verify that $S^+_t$ is monotone and $S^-_t$ is antimonotone. The
set $C$ is an antichain: if
\[
(x,\overline x)\preceq(y,\overline y),
\]
then $x\preceq y$ and $\overline x\preceq\overline y$, so $x=y$.
Moreover, $D$ is downward closed, and every point of $C$ is maximal in
$D$. Indeed, if $c\in C$ and $c\preceq z\in D$, then $z\preceq c'$
for some $c'\in C$. The antichain property gives $c=c'$, and hence $z=c$.

Now let $z\preceq z'$. If $z\in D\setminus C$, then $S^+_t(z)=0$, so
\[
S^+_t(z)\le S^+_t(z').
\]
If $z\notin D$, then $z'\notin D$, and both values are $1$. Finally,
if $z\in C$, then either $z'=z$ or $z'\notin D$, so the same inequality
holds. Thus $S^+_t$ is monotone. The function $1-S^-_t$ has the same
form, with values $1-\beta^-_t(x)$ on $C$, $0$ on $D\setminus C$, and
$1$ outside $D$. The same argument shows that $1-S^-_t$ is monotone,
so $S^-_t$ is antimonotone.

We now show that $f_s$ maps $G_i$ to itself. For every input
$u=(a,b,X,Y)$, each coordinate of the $X$-block of $f_s(u)$ lies between
$ah$ and $ah+k$. Since $0\le a\le k_i$,
\[
0\le ah\le ah+k\le k_i h+k=(k_i+1)h-1.
\]
Thus this block lies within the prescribed grid, whose coordinate widths
are $(k_i+1)h$. The same argument applies to the $Y$-block. Since
$S^+_b,S^-_a$ take values in $\{0,1\}$, the scalar outputs satisfy
\[
0\le\min\{k_i,b+S^+_b(Z_b(Y))\}\le k_i
\]
and
\[
0\le\max\{0,a-S^-_a(Z_a(X))\}\le a\le k_i.
\]
Hence $f_s(u)\in G_i$.

To prove monotonicity, let
\[
u=(a,b,X,Y)\preceq
\widetilde u=(\widetilde a,\widetilde b,\widetilde X,\widetilde Y)
\]
be two points in $G_i$. We compare their outputs block by block.

For the $X$-block, suppose first that $a=\widetilde a$. Since
$X\preceq\widetilde X$, monotonicity of $Z_a$ and $f'_{s_a}$ gives
\[
ah\mathbbm{1}+f'_{s_a}(Z_a(X))
\preceq
ah\mathbbm{1}+f'_{s_a}(Z_a(\widetilde X)).
\]
If instead $a<\widetilde a$, then every coordinate of $f_s(u)_X$ is
at most $ah+k$, whereas every coordinate of $f_s(\widetilde u)_X$ is
at least $\widetilde a h$. Since
\[
ah+k<(a+1)h\le\widetilde a h,
\]
we again obtain
\[
f_s(u)_X\preceq f_s(\widetilde u)_X.
\]
The same argument, using $b\le\widetilde b$ and
$Y\preceq\widetilde Y$, proves the required inequality for the $Y$-block.

For the $a$-coordinate, if $b=\widetilde b$, then monotonicity of $Z_b$
and $S^+_b$, followed by monotonicity of clipping, gives
\[
f_s(u)_a\le f_s(\widetilde u)_a.
\]
If $b<\widetilde b$, then
\[
f_s(u)_a\le b+1\le\widetilde b
\le f_s(\widetilde u)_a,
\]
where the last inequality follows from $S^+_{\widetilde b}\ge0$ and
$\widetilde b\le k_i$.

Finally, consider the $b$-coordinate. If $a=\widetilde a$, then
\[
Z_a(X)\preceq Z_a(\widetilde X).
\]
Since $S^-_a$ is antimonotone,
\[
a-S^-_a(Z_a(X))
\le
a-S^-_a(Z_a(\widetilde X)).
\]
Clipping below at zero preserves this inequality. If $a<\widetilde a$,
then
\[
f_s(u)_b\le a\le\widetilde a-1
\le f_s(\widetilde u)_b,
\]
where the last inequality follows from $S^-_{\widetilde a}\le1$.

Thus every output coordinate of $f_s(u)$ is at most the corresponding
coordinate of $f_s(\widetilde u)$. Therefore $f_s$ is monotone.
\end{proof}

\section{Proof of Lemma~\ref{lem:minjective}}
\label{app:minjective}

\begin{proof}
Let
\[
u=(a,b,X,Y)
\qquad\text{and}\qquad
\widetilde u=(\widetilde a,\widetilde b,\widetilde X,\widetilde Y)
\]
be fixed points of $f_s$ such that neither $m_s(u)$ nor
$m_s(\widetilde u)$ is a violation. Suppose that
\[
m_s(u)=m_s(\widetilde u)=v.
\]
We show that $u=\widetilde u$.

By the definition of $m_s$, a proper fixed point of the slice $s$ is
returned only in the case where
\[
a=b,\qquad
Z_a(X)=(x,\overline x),\qquad
Z_a(Y)=(x,\overline x)
\]
for some fixed point $x$ of $f_{s_a}$, with
\[
m_{s_a}(x)=v.
\]
Since $v$ belongs to the slice $s_a$, its $i$th coordinate is $a$. Thus
\[
a=b=v_i.
\]
Applying the same argument to $\widetilde u$, we obtain a fixed point
$\widetilde x$ of $f_{s_{\widetilde a}}$ such that
\[
\widetilde a=\widetilde b=v_i,
\qquad
Z_{\widetilde a}(\widetilde X)
=Z_{\widetilde a}(\widetilde Y)
=(\widetilde x,\overline{\widetilde x}),
\qquad
m_{s_{\widetilde a}}(\widetilde x)=v.
\]
In particular, $a=\widetilde a$, so $x$ and $\widetilde x$ are fixed
points of the same child function $f_{s_a}$. The inductive injectivity
property states that two distinct fixed points of $f_{s_a}$ cannot both
be mapped to the proper slice fixed point $v$. Therefore
\[
x=\widetilde x.
\]

It remains to show that the translated blocks agree. Since $u$ is a
fixed point of $f_s$,
\[
X=a(k+1)\mathbbm{1}+f'_{s_a}(Z_a(X)).
\]
Using $Z_a(X)=(x,\overline x)$ and $f_{s_a}(x)=x$, we have
\[
f'_{s_a}(x,\overline x)
=\bigl(f_{s_a}(x),\overline{f_{s_a}(x)}\bigr)
=(x,\overline x).
\]
Consequently,
\[
X=a(k+1)\mathbbm{1}+(x,\overline x).
\]
The same argument applies to $Y$, since $b=a$ and
$Z_a(Y)=(x,\overline x)$. Hence
\[
X=Y=a(k+1)\mathbbm{1}+(x,\overline x).
\]
Applying these identities to $\widetilde u$, and using
$\widetilde a=a$ and $\widetilde x=x$, gives
\[
\widetilde X=\widetilde Y
=a(k+1)\mathbbm{1}+(x,\overline x).
\]
Thus all coordinates of $u$ and $\widetilde u$ agree, so
$u=\widetilde u$.

Therefore two distinct fixed points of $f_s$ whose decoded outputs are
not violations must have distinct decoded outputs.
\end{proof}

\section{Proof of Lemma~\ref{lem:tarski-circuits}}
\label{app:tarski-circuits}

\begin{proof}
For each $i\in\{1,\dots,d\}$, we construct Boolean circuits $F_i$ and
$M_i$ such that
\[
F_i(s,x)=f_s(x)
\qquad\text{and}\qquad
M_i(s,x)=m_s(x)
\]
for every $i$-slice $s$, with the second equality required whenever $x$ is
a fixed point of $f_s$. The circuit $M_i$ may return an arbitrary output
on other inputs. We encode its output with a tag distinguishing a proper
slice fixed point from each type of violation.

Let $q_i$ be the number of dimensions of $G_i$, and let $W_i$ be its
maximum width. The bounds established in the main body of the paper give
\[
q_i=O(4^i)
\qquad\text{and}\qquad
W_i+1\le\prod_{j=1}^{i}(k_j+2).
\]
Consequently, the number of bits needed to encode one coordinate of $G_i$
is
\[
O\!\left(1+\sum_{j=1}^{i}\log(k_j+2)\right)=O(N),
\]
since the widths $k_j$ are included in the encoding of $\mathcal{O}$.
Thus a point of $G_i$ can be encoded using $O(4^iN)$ bits. A slice of
$\mathcal{O}$, a point of its original grid, or a violation of
$\mathcal{O}$ can each be encoded using polynomially many bits in $N$.

In the base case, the definitions of $f_s$ and $m_s$ require only a
constant number of evaluations of $D_1$, together with arithmetic,
comparisons, and selection operations on polynomially many bits. Hence
$F_1$ and $M_1$ have size polynomial in $N$, and can be constructed in
polynomial time.

Now suppose that $F_{i-1}$ and $M_{i-1}$ have been constructed. We first
apply the grid-equalization construction from Lemma~\ref{lem:squaregrid}.
This adds coordinate-wise clipping before evaluating $F_{i-1}$, so it
requires one copy of that circuit and additional circuitry of size
$O(4^{i-1}\poly(N))$. Since equalization preserves the fixed-point set,
the decoder continues to work on every fixed point. If necessary, we also
clip its input to the original grid before evaluating $M_{i-1}$; this
does not change its value at any fixed point.

We next construct $F_i$. Given $s$ and $(a,b,X,Y)$, the child slices
$s_a,s_b$ are obtained by inserting $a,b$ into the $i$th coordinate of
$s$. Computing the two mirrored output blocks requires four evaluations
of the equalized child function: two for each mirrored instance.

The scalar outputs require evaluating $S^+_b$ and $S^-_a$. Membership in
the sets $C,D$ can be tested coordinate-wise, since
\[
(x,y)\in C
\quad\Longleftrightarrow\quad
x_j+y_j=k\ \text{for every }j,
\]
and
\[
(x,y)\in D
\quad\Longleftrightarrow\quad
x_j+y_j\le k\ \text{for every }j.
\]
To evaluate either $\beta^+_t(x)$ or $\beta^-_t(x)$, we evaluate the child
function to test whether $x$ is a fixed point, evaluate the child decoder,
and, when its output is a proper slice fixed point, evaluate $D_i$ at
that output. The remaining cases are selected according to the definition
of the relevant bit function.

Thus the construction of $F_i$ uses only a constant number of copies of
the child function and decoder circuits, together with a constant number
of copies of the original direction circuit $D_i$. The additional
operations---reflection, translation, clipping, membership tests,
comparisons, and selection---can all be implemented by circuitry of size
\[
O(4^i\poly(N)).
\]

To construct $M_i$, compute the child points $x,y,p,q$ encoded by $X,Y$,
and evaluate the child decoder at
\[
(s_a,x),\quad(s_a,y),\quad(s_b,p),\quad(s_b,q).
\]
When the input is a fixed point of $f_s$, these are valid child fixed
points, so the inductive guarantees apply. Comparisons and selection
operations then implement the case distinction in the definition of
$m_s$: return a child violation, construct an \solnref{OV1} or
\solnref{OV2} violation, or return the proper slice fixed point. This
uses four copies of the child decoder and additional circuitry of size
$O(4^i\poly(N))$.

In particular, neither construction enumerates the child slices. The
values $a,b$ are supplied as inputs to copies of the uniform child
circuits.

Let $A_i$ be the combined size of $F_i$ and $M_i$. The preceding
constructions give an absolute constant $c$ and a fixed polynomial $p$
such that
\[
A_1\le p(N),
\qquad
A_i\le cA_{i-1}+4^ip(N).
\]
Expanding this recurrence yields
\[
A_i
\le c^{i-1}p(N)
+p(N)\sum_{j=2}^{i}c^{i-j}4^j
=\poly(N)\,2^{O(i)}.
\]

All of the circuit operations described above are explicit, and the
circuits can be generated in time polynomial in $N$ and their total
size. Accounting for all $d$ levels therefore gives construction time
\[
\poly(N)\,2^{O(d)}.
\]
Finally, fixing the slice input of $F_d$ and $M_d$ to the unique all-free
$d$-slice produces the circuits for $f$ and $m$, respectively, within
the same bound.
\end{proof}

\section{Cryptographic lower bounds}
\label{app:crypto}

In this section we obtain time lower bounds for \Tarski under
cryptographic assumptions. We first reduce two line-following problems
to \UEOPL, keeping track of the number of potential bits, and then
apply Theorem~\ref{thm:main}. Throughout this section, logarithms are
to base two, and
$\lambda$ denotes the cryptographic security parameter. We continue to
write $N$ for the encoding size of a \UEOPL instance, $n$ for the number
of bits in its potential, and $M$ for the encoding size of a \Tarski
instance.

The lower bounds below apply to randomized algorithms that succeed with
probability at least $2/3$ on every input. The cryptographic hardness
statements are against non-uniform circuits. A success probability is
\emph{negligible} in $\lambda$ if, for every constant $c>0$, it is at most
$\lambda^{-c}$ for all sufficiently large $\lambda$. All distributions
of source instances used below can be sampled in time polynomial in
$\lambda$.

\subsection{From \rSVL to \UEOPL to \Tarski}
\label{sec:rsvl-ueopl-tarski}

We first reduce relaxed sink of verifiable line (\rSVL) to unique
forward end of metered line (\UFEOML), and then reduce \UFEOML to
\UEOPL. The second reduction can also be applied directly to hard
\UFEOML instances. We use the definition of \rSVL from~\cite{EFKP20},
with indices starting at zero.

\begin{definition}[\rSVL and \SVL]
\label{def:rsvl}
Given an integer $L\ge 1$, a point
$v_0\in\Set{0,1}^m$, and Boolean circuits
\[
S:\Set{0,1}^m\to\Set{0,1}^m,
\qquad
W:\Set{0,1}^m\times[L]\to\Set{0,1},
\]
with the promise that $W(S^i(v_0),i)=1$ for every $i\in[L]$, find
one of the following:
\begin{enumerate}[label=(R\arabic*), wide=0pt]
\item \solnlabel{R1} A point $v$ such that $W(v,L)=1$.
\item \solnlabel{R2} A pair $(v,i)$ such that $i\in[L]$, $W(v,i)=1$, and
$v\ne S^i(v_0)$.
\end{enumerate}
In sink of verifiable line (\SVL), the promise is strengthened to
\[
W(v,i)=1\quad\Longleftrightarrow\quad v=S^i(v_0)
\]
for every $v\in\Set{0,1}^m$ and $i\in[L]$, and the task is to
find a solution of type \solnref{R1}.
\end{definition}

As in the rest of the paper, $[L]=\{0,1,\dots,L\}$. Thus our indices
start at zero; shifting the indices in a source that starts at one
changes the length by at most one. A solution of type \solnref{R2} is
called a \emph{false positive}. In particular, a \SVL instance is a
\rSVL instance with no false positives.

\begin{definition}[\UFEOML]
\label{def:ufeoml}
Given Boolean circuits
\[
S:\Set{0,1}^m\to\Set{0,1}^m,
\qquad
V:\Set{0,1}^m\to\Set{0,1,\dots,2^b-1},
\]
and a point $v_0$ such that $S(v_0)\ne v_0$ and $V(v_0)=0$, find
one of the following:
\begin{enumerate}[label=(F\arabic*), wide=0pt]
\item \solnlabel{F1} A point $v$ such that $S(v)\ne v$ and $S(S(v))=S(v)$.
\item \solnlabel{F2} A point $v$ such that $S(v)\ne v$ and
$V(S(v))\ne V(v)+1$.
\item \solnlabel{F3} Two distinct points $u,v$ such that $S(u)\ne u$, $S(v)\ne v$,
and $V(u)=V(v)$.
\end{enumerate}
\end{definition}

The circuit $V$ is called the \emph{meter}. The three types of solution
are a sink, a meter violation, and a meter collision, respectively.
This is the problem called UniqueForwardEOPL$+1$ in~\cite{FGMS20},
with its first solution type split into two cases.

\paragraph{\bf From \rSVL to \UFEOML.}

The following reduction is the one in~\cite[Appendix~C]{Choudhuri26},
with the index length recorded explicitly.

\begin{lemma}
\label{lem:rsvl-ufeoml}
Let $\mathcal{I}$ be a \rSVL instance satisfying the promise, with
encoding size $K$ and length $L$, and put
$b=\lceil\log(L+1)\rceil$. In time
$\poly(K,b)$ we can construct a \UFEOML instance of size
$\poly(K,b)$ whose meter uses $b$ bits. Given any solution of this
instance, we can in time $\poly(K,b)$ output an \rSVL solution with
probability at least $1/2$.
\end{lemma}
\begin{proof}
Write $\mathcal{I}=(S,W,L,v_0)$. The new vertices are pairs $(v,i)$,
where $v\in\Set{0,1}^m$ and $i$ is a $b$-bit integer. Define
\[
S'(v,i)=
\begin{cases}
(S(v),i+1) & \text{if }i<L\text{ and }W(v,i)=1,\\
(v,i) & \text{otherwise},
\end{cases}
\qquad
V'(v,i)=i.
\]
The designated starting point is $(v_0,0)$. The promise on $W$ ensures
that this point has a nontrivial successor, and $V'(v_0,0)=0$.
The circuits have the claimed size. In particular, no enumeration of
the $L$ steps is required.

There are no meter violations, since the meter increases by one on
every nontrivial successor step. Suppose that $(v,i)$ is a sink, and
write $w=S(v)$. Then $i<L$, $W(v,i)=1$, and either $i+1=L$ or
$W(w,i+1)=0$. If $i+1=L$ and $W(w,L)=1$, we return $w$.
In every other case $W(w,i+1)=0$. The promise then implies that
$v\ne S^i(v_0)$, so $(v,i)$ is a false positive.

Finally, a meter collision consists of distinct pairs $(u,i)$ and
$(v,i)$ with $W(u,i)=W(v,i)=1$. Since $u\ne v$, at most one of these
points equals $S^i(v_0)$. We return one of $(u,i)$ and $(v,i)$ uniformly
at random. This is a false positive with probability at least $1/2$.
\end{proof}

\paragraph{\bf From \UFEOML to \UEOPL.}

We use the reversible pebbling reduction
of~\cite[Lemma~15 and Appendix~B.2]{FGMS20}. This builds on the
\SVL-to-end-of-line reduction of Bitansky, Paneth and Rosen~\cite{BPR15}
and the metering construction of Hub{\'a}{\v c}ek and Yogev~\cite{HY17}.
The underlying reversible pebbling technique is due to Bennett~\cite{Ben89};
its use in end-of-line constructions already appears in Abbott, Kane
and Valiant~\cite{AKV04}.
The potential records progress through the pebbling schedule.
We record the bounds in terms of the number of bits in the meter.

\begin{lemma}
\label{lem:ufeoml-ueopl}
Let $\mathcal{F}$ be a \UFEOML instance of encoding size $K$, with
$m$-bit vertices and a $b$-bit meter. In time $\poly(K,b)$ we can
construct a \UEOPL instance of encoding size
\[
N=\poly(K,b)
\]
whose vertices use $O((m+b)(b+1))$ bits and whose potential uses
$n=O(b+1)$ bits. Every solution of the \UEOPL instance can be mapped
in time $\poly(K,b)$ to a solution of $\mathcal{F}$.
\end{lemma}
\begin{proof}
Write $\mathcal{F}=(S,V,v_0)$. If $v_0$ is already a solution of
type \solnref{F1} or \solnref{F2}, we output a fixed one-edge
\UEOPL instance and map its solution to $v_0$. Otherwise, we use
the construction from~\cite[Lemma~15]{FGMS20}, which implements
a reversible walk using a fixed recursive pebbling schedule. A pebble
stores a vertex of $\mathcal{F}$ together with its meter value. The
schedule needs only $b+1$ pebbles: a sequence of steps on which the
meter increases by one has length less than $2^b$, and reaching a
larger meter value would already give a meter violation. Occupied
registers represent points $v$ with $S(v)\ne v$. If the simulation
attempts to move from such a point to a point fixed by $S$, then $v$
is a sink of the original instance and the simulation stops there.
The designated starting configuration has a nontrivial successor and
can be renamed $0^{m'}$, as required by
Definition~\ref{def:UEOPL}. We number the schedule positions from zero,
so its potential is also zero.

We record the size bounds in this construction. If $a_j$ is the number
of moves in the schedule with $j$ pebbles, its recursive definition
gives
\[
a_0=0,
\qquad
a_j=3a_{j-1}+1,
\qquad
a_j=(3^j-1)/2.
\]
The successor and predecessor circuits implement one move of this
schedule. The potential circuit computes the position in the schedule
from the occupied pebble indices, using the recurrence
in~\cite[Claim~6.6 and Algorithm~1]{HY17}. These computations follow
a recursion of depth $O(b)$ on integers with $O(b)$ bits. In particular,
they do not enumerate the schedule. A configuration uses
$O((m+b)(b+1))$ bits, and the circuits can be constructed in time
$\poly(K,b)$.

The potential is bounded by $3^{b+1}+O(1)$ on every configuration
retained by the reduction. Malformed configurations are isolated, with
potential zero, and unused pebble registers have a fixed canonical
encoding. Thus the potential uses $O(b+1)$ bits globally, including
away from the line starting at the designated source.

For completeness, we recall how the solution types are handled
in~\cite[Appendix~B.2]{FGMS20}. An endpoint of type \solnref{U1}
or \solnref{UV2} arises when a forward or backward pebbling move
cannot be performed. If placing a pebble fails, the point from which
we try to move gives a solution of type \solnref{F1} or
\solnref{F2}. If removing a pebble at $v$ fails because $S(u)\ne v$
for the preceding pebble $u$, we first check whether $u$ is a
solution of type \solnref{F1} or \solnref{F2}. If it is not, then
$S(u)$ and $v$ are distinct points with nontrivial successors and the
same meter, and give a solution of type \solnref{F3}. The complete schedule
cannot be reached without encountering a solution, since it would
require a meter value at least $2^b$.

The potential increases by one on every edge whose successor and
predecessor agree, so there are no violations of type \solnref{UV1}.
For a solution of type \solnref{UV3}, first check whether either
point is already a solution of type \solnref{U1}. Otherwise, the
unit potential increments imply that the two configurations have the
same potential. They therefore refer to the same step of the schedule,
with identical occupied pebble indices. Since unused registers have
canonical encodings, some occupied register differs, giving a meter
collision of type \solnref{F3}. All these checks take time
$\poly(K,b)$.
\end{proof}

Combining the two parts with Theorem~\ref{thm:main} gives the bound
that we will use in each application.

\begin{corollary}
\label{cor:rsvl-tarski}
Let $\mathcal{I}$ be a \rSVL instance satisfying the promise, with
encoding size $K$ and length $L$. There is a reduction to a violation-free \Tarski
instance of encoding size $M$ whose construction and decoding time,
as well as $M$, are bounded by
\[
\poly(K,\log(L+1))\cdot
2^{O(\sqrt{\log(L+1)})}.
\]
Every fixed point can be decoded to a solution of $\mathcal{I}$ with
probability at least $1/2$. For a \UFEOML instance of size $K$ with
a $b$-bit meter, the corresponding bound is
\[
\poly(K,b)\cdot 2^{O(\sqrt{b+1})},
\]
and the decoding is deterministic.
\end{corollary}
\begin{proof}
Apply Lemmas~\ref{lem:rsvl-ufeoml} and~\ref{lem:ufeoml-ueopl}, or
only the second lemma when starting from \UFEOML. We obtain a
\UEOPL instance with $N=\poly(K,b)$ and $n=O(b+1)$, where
$b=\lceil\log(L+1)\rceil$ in the \rSVL case. We may pad the
potential with leading zeros so that $n\ge 4$. Set
$r=2^{\lfloor\sqrt n\rfloor}$ in Theorem~\ref{thm:main}.
Then $r\le 2^{n-2}$ and the construction and decoding take time
\[
\poly(N,r)\cdot 2^{O(n/\log(r+1))}
=\poly(K,b)\cdot 2^{O(\sqrt{b+1})}.
\]
The encoding size of the resulting instance is bounded by the
construction time. Theorem~\ref{thm:main} guarantees that the map is
monotone even when the source instance has violations. The only
randomness in decoding is the choice in a meter collision in
Lemma~\ref{lem:rsvl-ufeoml}.
\end{proof}

\subsection{Repeated squaring and strong Fiat--Shamir}
\label{sec:crypto-efkp}

Ephraim, Freitag, Komargodski and Pass~\cite{EFKP20} construct
\rSVL instances whose hardness is proportional to their length. We
use the following form of their result. The $(D,B)$ repeated-squaring
assumption and the strong Fiat--Shamir assumption below are
Assumptions~3.4 and~3.7 of~\cite{EFKP20}, respectively. The latter
requires soundness also for public-coin proofs with an inefficient
verifier.

\begin{proposition}[{\cite[Theorem~7.8 and Corollary~7.9]{EFKP20}}]
\label{prop:efkp-source}
Let $B=B(\lambda)$ be efficiently describable and satisfy
\[
\lambda^c\le B\le 2^{\lambda^{1/3}}
\]
for a sufficiently large constant $c$. Assume the $(D,B)$
repeated-squaring assumption for a polynomial $D$ satisfying
$D\ge\lambda^{d_0}$, where $d_0$ is the constant in the construction,
and the strong Fiat--Shamir assumption for the public-coin proofs
with up to $\lceil\log_\lambda B\rceil$ rounds.
There is a polynomial-time sampler of promised \rSVL instances with
encoding size $K=\poly(\lambda)$ and length $L=\Theta(B)$, and a
constant $c_0>0$, such that every circuit family of size at most
$c_0B$ solves the sampled instance with only negligible probability.
\end{proposition}

\begin{corollary}
\label{cor:efkp-tarski}
Under the assumptions of Proposition~\ref{prop:efkp-source} with
$B=2^{\lfloor(\log\lambda)^2\rfloor}$, \Tarski cannot be solved
in time $M^{o(\log M)}$, even on violation-free instances.
\end{corollary}
\begin{proof}
For these parameters $\log(L+1)=\Theta((\log\lambda)^2)$.
Corollary~\ref{cor:rsvl-tarski} therefore constructs a \Tarski
instance, and decodes any fixed point, in time $\poly(\lambda)$.
Its encoding size satisfies $M=\poly(\lambda)$.

Suppose that a \Tarski algorithm ran in time $M^{o(\log M)}$.
Composing it with the reduction and decoder would solve the source
\rSVL instance with probability at least $1/3$ in time
\[
\poly(\lambda)+2^{o((\log\lambda)^2)}
=2^{o((\log\lambda)^2)}.
\]
The resulting circuit size, including the polynomial overhead of
simulating this computation by a circuit, is $B^{o(1)}<c_0B$ for
all sufficiently large $\lambda$. This contradicts
Proposition~\ref{prop:efkp-source}.
\end{proof}

This instantiation requires $\Theta(\log\lambda)$ rounds in the
strong Fiat--Shamir assumption. A different instantiation gives
the Weak ETH for \UEOPL used in the main body of the paper.

\begin{corollary}
\label{cor:efkp-weak-eth}
Under the assumptions of Proposition~\ref{prop:efkp-source} with
$B=2^{\lfloor\lambda^{1/3}\rfloor}$, the Weak ETH for \UEOPL
in Conjecture~\ref{conj:weak-eth} holds. In fact, for some constant
$c$, the instances with $N\le n^c$ cannot be solved in time
$\poly(N)\cdot 2^{o(n)}$.
\end{corollary}
\begin{proof}
Apply Lemmas~\ref{lem:rsvl-ufeoml} and~\ref{lem:ufeoml-ueopl} to
the distribution in Proposition~\ref{prop:efkp-source}. The resulting
\UEOPL instances have $N=\poly(\lambda)$ and
$n=O(\lambda^{1/3})$. By padding the potential with leading zeros,
we may take $n=\Theta(\lambda^{1/3})$, and hence $N\le n^c$ for
some fixed $c$. An algorithm with the stated running time would,
after construction and decoding, solve the \rSVL instance in time
\[
\poly(\lambda)\cdot 2^{o(\lambda^{1/3})}
=2^{o(\lambda^{1/3})}
\]
with probability at least $1/3$. This again gives circuits smaller
than $c_0B$, contradicting Proposition~\ref{prop:efkp-source}.
\end{proof}

Corollary~\ref{cor:efkp-weak-eth} uses the cryptographic assumptions
at the larger value $B=2^{\lfloor\lambda^{1/3}\rfloor}$. It therefore
requires the strong Fiat--Shamir assumption for
$\Theta(\lambda^{1/3}/\log\lambda)$ rounds, whereas
Corollary~\ref{cor:efkp-tarski} uses $\Theta(\log\lambda)$ rounds.

\subsection{Functional encryption and injective one-way functions}
\label{sec:crypto-ks}

Komargodski and Segev~\cite{KS17} construct hard \SVL instances
from private-key functional encryption and injective one-way functions.
We use the following quantitative form of their result. Selective
security includes both message privacy and function privacy, as
in~\cite[Definition~2.7]{KS17}.

\begin{proposition}[{\cite[Theorem~4.10 and its proof]{KS17}}]
\label{prop:ks-source}
Fix a constant $a>1$ and put $Q(\lambda)=2^{(\log\lambda)^a}$.
Assume a single-input private-key functional encryption scheme for
polynomial-size circuits that is selectively secure for time
$Q(\lambda)^2$, at most $Q(\lambda)^2$ key queries, and at most
$Q(\lambda)$ encryption queries, with distinguishing advantage at
most $2^{-(\log\lambda)^{3a/2}}$. Let $\delta\in(0,1)$ be the
constant supplied by~\cite[Corollary~4.3]{KS17} for these parameters.
Assume also injective one-way functions that are secure against time $2^{\kappa^\rho}$ with
inversion probability at most $2^{-\kappa^\rho}$ at security
parameter $\kappa$, where
\[
\frac{1}{1+\delta/3}<\rho<1.
\]
There is a constant $\eta>0$ and a
polynomial-time sampler of \SVL instances with
\[
K=\poly(\lambda),
\qquad
\log(L+1)=\Theta((\log\lambda)^{1+\delta/3}),
\]
such that every circuit family of size at most
$2^{(\log\lambda)^{1+\eta}}$ solves the sampled instance with
only negligible probability.
\end{proposition}

Here $a=1/\eps$ in~\cite{KS17}. To check the one-way function
parameters, choose $\sigma$ with $1/(1+\delta/3)<\sigma<\rho$.
The assumed security at parameter $\kappa$ then implies
the $(2^{2\kappa^\sigma},2^{-2\kappa^\sigma})$ security used in
their proof, for all sufficiently large $\kappa$. The smallest
one-way function input in that proof has length
$\Theta((\log\lambda)^{1+\delta/3})$. Since
$\sigma(1+\delta/3)>1$, its security bound still exceeds
$2^{(\log\lambda)^{1+\eta}}$ when
$0<\eta<\sigma(1+\delta/3)-1$.
The other security terms and the sum over the line positions are
controlled by $\sigma(1+\delta)>1+\delta/3$ and
$2>1+\delta/3$. Thus the bounds proved in~\cite[Theorem~4.10]{KS17}
give the stated quasipolynomial adversary size and negligible
success probability.

\begin{corollary}
\label{cor:ks-tarski}
Under the assumptions of Proposition~\ref{prop:ks-source}, there
is a constant $\eta>0$ such that \Tarski cannot be solved in time
$2^{o((\log M)^{1+\eta})}$, even on violation-free instances.
\end{corollary}
\begin{proof}
Regard the \SVL instances as \rSVL instances, and apply
Corollary~\ref{cor:rsvl-tarski}. Since
$\delta<1$, the construction and decoding time is at most
\[
\poly(\lambda)\cdot
2^{O((\log\lambda)^{(1+\delta/3)/2})}
=\poly(\lambda),
\]
and $M=\poly(\lambda)$. A \Tarski algorithm with the stated
running time would consequently solve the source instance with
probability at least $1/3$ in time
\[
\poly(\lambda)+2^{o((\log\lambda)^{1+\eta})}
=2^{o((\log\lambda)^{1+\eta})}.
\]
Converting this computation into a circuit preserves the
little-$o$ bound in the exponent, contradicting
Proposition~\ref{prop:ks-source}.
\end{proof}

\subsection{LWE and iterated squaring}
\label{sec:crypto-bch}

Bitansky et al.~\cite{BCHKLPR22} establish hardness of \UFEOML from
polynomially hard LWE and polynomial-time hardness of iterated
squaring. Here LWE denotes the assumption used
in~\cite[Assumption~2.10 and Corollary~4.10]{BCHKLPR22}. For iterated
squaring, sample a group $\mathcal{G}_\lambda$ with unknown order
and a uniform element $g$ as in~\cite[Definition~4.9]{BCHKLPR22}. For
a specified number $T$ of squarings, the computational task is to output
$g^{2^T}$. The assumption is that no circuit family of size
$\poly(\lambda)$ succeeds with non-negligible probability. One
may use the RSA groups and sampling procedure
in~\cite[Assumptions~4.3 and~4.8]{BCHKLPR22}.

\begin{proposition}[{\cite[Corollary~B.4]{BCHKLPR22}}]
\label{prop:bch-source}
Let $T=2^t$, where $t$ is polynomially bounded in $\lambda$ and
efficiently computable. Assume polynomially hard LWE and
polynomial-time hardness of iterated squaring for $T$ squarings.
There is a polynomial-time sampler of \UFEOML instances of
encoding size $K=\poly(\lambda,t)$ with a meter using
$b=O(t+1)$ bits, such that every polynomial-size circuit family
solves the sampled instance with only negligible probability.
\end{proposition}

The meter bound follows from the construction
in~\cite[Appendix~B]{BCHKLPR22}, which has length $L=3^t$ and uses
the index in a restricted depth-first traversal as its meter.
Whenever $S(v)\ne v$, both $V(v)$ and $V(S(v))$ lie in $[L]$.
Replace the paper's padded meter output by
\[
\widehat V(v)=\min\{V(v),L\},
\]
encoded using $\lceil\log(L+1)\rceil=O(t+1)$ bits. This preserves
the meter at every point $v$ with $S(v)\ne v$ and at its successor.
Solutions of
type \solnref{F1} depend only on $S$, and the meter values used in
types \solnref{F2} and \solnref{F3} are unchanged. Hence this
replacement preserves every solution and bounds the meter on all
inputs, including malformed encodings.

\begin{corollary}
\label{cor:bch-tarski}
Assume polynomially hard LWE and polynomial-time hardness of
iterated squaring at
$T=2^{\lceil(\log\lambda)^2\rceil}$. Then \Tarski has no
polynomial-time algorithm, even on violation-free instances.
\end{corollary}
\begin{proof}
Proposition~\ref{prop:bch-source} gives $K=\poly(\lambda)$
and $b=O((\log\lambda)^2)$. Apply the \UFEOML part of
Corollary~\ref{cor:rsvl-tarski}. The construction, the size
$M$ of the resulting instance, and the decoding time are all
bounded by
\[
\poly(\lambda)\cdot 2^{O(\log\lambda)}
=\poly(\lambda).
\]
A polynomial-time \Tarski algorithm would therefore give a
polynomial-time solver for the source distribution, with success
probability at least $2/3$, contradicting
Proposition~\ref{prop:bch-source}.
\end{proof}

The specified value of $T$ is part of the hypothesis of this corollary.
The existential assertion that iterated squaring is hard for
\emph{some} efficiently describable $T$, as in
\cite[Assumption~4.3]{BCHKLPR22}, does not by itself ensure
$\log T=O((\log\lambda)^2)$.

\subsection{Polynomial SIS and planted search}
\label{sec:crypto-sis-planted}

Choudhuri~\cite{Choudhuri26} gives a hardness result from
polynomially hard SIS together with an exponentially hard planted
search problem. We state both assumptions, since the norm bound
in the SIS assumption and the independent search assumption are
relevant to this application.

\paragraph{\bf The SIS assumption.}
Fix a sufficiently large absolute constant $C$. For every
polynomially bounded $\kappa=\kappa(\lambda)\ge\lambda^{\Omega(1)}$
with $\kappa\ge2$, put
\[
\beta=2^{C\lceil\log\kappa\rceil^3}.
\]
For every polynomially bounded $\mu=\mu(\lambda)$ and every
power of two $q$ satisfying
\[
\kappa\sqrt\mu\,\beta\le q,
\qquad
q/\beta=\poly(\kappa),
\]
assume that, given a uniform matrix
$A\in\ints_q^{\kappa\times\mu}$, no polynomial-size circuit
family finds $z\in\ints^\mu$ satisfying
\[
Az=0\pmod q,
\qquad
0<\|z\|_\infty\le\beta
\]
with non-negligible probability. This is the assumption
$\mathrm{SIS}_{\mathrm{qp}}$
of~\cite[Assumption~3.6]{Choudhuri26}. In particular, the norm
bound is quasipolynomial, although the assumed adversary size is
only polynomial.

\paragraph{\bf The planted-search assumption.}
There is a polynomial-time sampler that, on input $1^\ell$,
outputs $(z,w)$ with $w\in\Set{0,1}^\ell$ and $R(z,w)=1$, for
a polynomial-time relation $R$. Assume that there is a constant
$\gamma>0$ such that, for all sufficiently large $\ell$, every
circuit of size at most $2^{\gamma\ell}$, given $z$, outputs
some $w'$ with $R(z,w')=1$ with probability at most
$2^{-\gamma\ell}$. This is~\cite[Assumption~7.1]{Choudhuri26}.
For example, a one-way function with these bounds on circuit size
and inversion probability gives the assumption by sampling $w$
uniformly, setting $z=g(w)$, and taking $R(z,w')=1$ if $g(w')=z$.

\begin{proposition}[{\cite[Theorem~4.5, Lemma~7.2 and Corollary~7.3]{Choudhuri26}}]
\label{prop:sis-planted-source}
Under the SIS and planted-search assumptions above, there is a
polynomial-time sampler of promised \rSVL instances with
\[
K=\poly(\lambda),
\qquad
L=2^{\lceil(\log\lambda)^2\rceil},
\]
such that every polynomial-size circuit family solves the sampled
instance with only negligible probability.
\end{proposition}

This is the \rSVL hardness statement underlying the EOML
conclusion of~\cite[Corollary~7.3]{Choudhuri26}. Its construction
already uses height $H=\lceil(\log\lambda)^2\rceil$ and length
$L=2^H$. We use this construction with its stated parameters and
polynomial-size adversaries.

\begin{corollary}
\label{cor:sis-planted-tarski}
Under the assumptions of Proposition~\ref{prop:sis-planted-source},
\Tarski has no polynomial-time algorithm, even on violation-free
instances.
\end{corollary}
\begin{proof}
Apply Corollary~\ref{cor:rsvl-tarski} to the distribution in
Proposition~\ref{prop:sis-planted-source}. Since
$\log(L+1)=\Theta((\log\lambda)^2)$, the construction and
decoding take time $\poly(\lambda)$ and the resulting encoding
size is $M=\poly(\lambda)$. A polynomial-time \Tarski algorithm
would therefore solve the source \rSVL instance in polynomial
time with probability at least $1/3$, contradicting
Proposition~\ref{prop:sis-planted-source}.
\end{proof}

Among the instantiations in this section, only
Corollary~\ref{cor:efkp-weak-eth} establishes the Weak ETH for
\UEOPL. The other applications above give direct lower bounds
for \Tarski: their source bounds do not supply both an encoding
size polynomial in the number of potential bits and the hardness
required by Conjecture~\ref{conj:weak-eth}.

\end{document}